\documentclass{lmcs}
\pdfoutput=1

\usepackage{lastpage}
\lmcsdoi{16}{1}{26}
\lmcsheading{}{\pageref{LastPage}}{}{}%
{Nov.~28,~2017}{Feb.~25,~2020}{}

\newif\ifignore 
\ignorefalse

\newcommand{\hide}[1]{}

\newcommand{\auxproof}[1]{
\ifignore\mbox{}\newline
\textbf{PROOF:} \dotfill\newline
{\it #1}\mbox{}\newline
\textbf{ENDPROOF}\dotfill
\fi}

\usepackage{amsmath}
\usepackage{amssymb}
\usepackage{amstext}
\usepackage{mathrsfs}
\usepackage{mathtools}
\usepackage{stmaryrd}
\usepackage{xspace}
\usepackage{cmll}

\usepackage{url}
\usepackage{fancybox}
\usepackage{nicefrac}
\usepackage{listings} 

\usepackage{xypic}
\usepackage[all]{xy}
\xyoption{all}
\xyoption{2cell}
\UseTwocells
\xyoption{tips}
\SelectTips{cm}{}  

\theoremstyle{plain}

\newtheorem{lemma}[thm]{Lemma}
\newtheorem{proposition}[thm]{Proposition}
\newtheorem{corollary}[thm]{Corollary}

\theoremstyle{definition}
\newtheorem{definition}[thm]{Definition}
\newtheorem{example}[thm]{Example}

\newenvironment{myproof}[1][Proof]%
   {\begin{proof}%
   }%
   {\end{proof}%
   }
\newenvironment{appendixproof}[1]
   {\begin{proof}[Proof #1]%
   }%
   {\end{proof}%
   }

\newcommand{\QEDbox}{\qedhere}
\newcommand{\QED}{\qedhere}

\makeatletter
\newcommand{\mathoverlap}[2]{\mathpalette\mathoverlap@{{#1}{#2}}}
\newcommand{\mathoverlap@}[2]{\mathoverlap@@{#1}#2}
\newcommand{\mathoverlap@@}[3]{\ooalign{$\m@th#1#2$\crcr\hidewidth$\m@th#1#3$\hidewidth}}
\makeatother

\DeclareSymbolFont{T1op}{T1}{cmr}{m}{n}
\SetSymbolFont{T1op}{bold}{T1}{cmr}{bx}{n}
\DeclareMathSymbol{\mathguilsinglleft}{\mathopen}{T1op}{'016}
\DeclareMathSymbol{\mathguilsinglright}{\mathclose}{T1op}{'017}

\newcommand{\idmap}[1][]{\ensuremath{\mathrm{id}_{#1}}}
\newcommand{\after}{\mathrel{\circ}}
\newcommand{\evmap}{\mathrm{ev}}

\newcommand{\tr}{\ensuremath{\mathrm{tr}}}

\newcommand{\upsum}[1]{\ensuremath{#1{\kern-.6ex}\uparrow}\xspace}
\newcommand{\downsum}[1]{\ensuremath{#1{\kern-.6ex}\downarrow}\xspace}

\newcommand{\tvd}{\ensuremath{\mathsf{tvd}}}
\newcommand{\kvd}{\ensuremath{\mathsf{kvd}}}
\newcommand{\trd}{\ensuremath{\mathsf{trd}}}
\newcommand{\ard}{\ensuremath{\mathsf{ard}}}
\newcommand{\spd}{\ensuremath{\mathsf{spd}}}
\newcommand{\vld}{\ensuremath{\mathsf{vld}}}

\newcommand{\orthogonal}{\mathrel{\bot}}
\newcommand{\set}[2]{\{#1\;|\;#2\}}
\newcommand{\setin}[3]{\{#1\in#2\;|\;#3\}}
\newcommand{\supp}{\mathrm{supp}}
\newcommand{\op}[1]{#1^{\mathrm{op}}}

\newcommand{\ex}[2]{\exists#1.\,#2}

\newcommand{\tuple}[1]{\langle#1\rangle}

\newcommand{\ket}[1]{\ensuremath{|{\kern.1em}#1{\kern.1em}\rangle}}
\newcommand{\bigket}[1]{\ensuremath{\big|{\kern.1em}#1{\kern.1em}\big\rangle}}
\newcommand{\bra}[1]{\langle\,#1\,|}

\newcommand{\andthen}{\mathrel{\&}}

\newcommand{\C}{\mathbb{C}}

\newcommand{\opnorm}[1]{\|#1\|_{\mathrm{op}}}

\newcommand{\distributionsymbol}{\mathcal{D}}

\newcommand{\Dst}{\distributionsymbol}

\newcommand{\UF}{\ensuremath{\mathcal{U}{\kern-.75ex}\mathcal{F}}}

\newcommand{\Cat}[1]{\ensuremath{\mathbf{#1}}\xspace}
\newcommand{\cat}[1]{\Cat{#1}}
\newcommand{\Kl}{\mathcal{K}{\kern-.4ex}\ell}
\newcommand{\Klf}{\Kl_{\mathrm{fin}}}
\newcommand{\EM}{\mathcal{E}{\kern-.4ex}\mathcal{M}}

\newcommand{\Sets}{\Cat{Sets}}

\newcommand{\Met}{\Cat{Met}}

\newcommand{\NNO}{\mathbb{N}}

\newcommand{\R}{\mathbb{R}}

\newcommand{\nnR}{\mathbb{R}_{\geq 0}}

\newcommand{\EMod}{\Cat{EMod}}
\newcommand{\AEMod}{\Cat{AEMod}}

\newcommand{\wEMod}{\ensuremath{\omega\text{-}\EMod}}
\newcommand{\DcEMod}{\Cat{DcEMod}}
\newcommand{\Conv}{\Cat{Conv}}
\newcommand{\ConvMet}{\Cat{ConvMet}}
\newcommand{\ConvCMet}{\Cat{ConvCMet}}
\newcommand{\vNA}{\Cat{vNA}}

\newcommand{\Hom}{\ensuremath{\mathrm{Hom}}}
\newcommand{\Pred}{\ensuremath{\mathrm{Pred}}}
\newcommand{\Stat}{\ensuremath{\mathrm{Stat}}}

\newcommand{\Ef}{\ensuremath{\mathcal{E}{\kern-.5ex}f}}
\newcommand{\intd}{{\kern.2em}\mathrm{d}{\kern.03em}}
\newcommand{\indic}[1]{\mathbf{1}_{#1}}

\newcommand{\bigovee}{\mathop{\vphantom{\sum}\mathchoice%
        {\vcenter{\hbox{\huge $\ovee$}}}%
        {\vcenter{\hbox{\Large $\ovee$}}}%
        {\ovee}{\ovee}}\displaylimits}

\newcommand{\OF}{\ensuremath{\mathcal{O}{\kern-.1em}\mathcal{F}}}
\newcommand{\Closed}{\ensuremath{\mathcal{C}{\kern-.05em}\ell}}

\makeatother
 \newdir{ >}{{}*!/-7.5pt/@{>}}
 \newdir{|>}{!/4.5pt/@{|}*:(1,-.2)@^{>}*:(1,+.2)@_{>}}
 \newdir{ |>}{{}*!/-3pt/@{|}*!/-7.5pt/:(1,-.2)@^{>}*!/-7.5pt/:(1,+.2)@_{>}}

\newcommand{\ie}{\textit{i.e.}\xspace}
\newcommand{\eg}{\textit{e.g.}\xspace}

\DeclareFixedFont{\ttb}{T1}{txtt}{bx}{n}{10} 
\DeclareFixedFont{\ttm}{T1}{txtt}{m}{n}{10}  
\usepackage{color}
\definecolor{deepblue}{rgb}{0,0,0.5}
\definecolor{deepred}{rgb}{0.6,0,0}
\definecolor{deepgreen}{rgb}{0,0.5,0}

\newcommand\pythonstyle{\lstset{
language=Python,
basicstyle=\ttm,
otherkeywords={self,>>>,...},             
keywordstyle=\ttb\color{deepblue},
emph={MyClass,__init__},          
emphstyle=\ttb\color{deepred},    
stringstyle=\color{deepgreen},
frame=tb,                         
showstringspaces=false            %
}}

\lstnewenvironment{python}[1][]
{
\pythonstyle
\lstset{#1}
}
{}

\newcommand\pythoninline[1]{{\pythonstyle\lstinline!#1!}}

\title{Distances between States and between Predicates}

\author{Bart Jacobs and Abraham Westerbaan}

\address{
Institute for Computing and Information Sciences, 
Radboud University Nijmegen, The Netherlands. 
}
\email{bart@cs.ru.nl, bram@westerbaan.name}
\urladdr{www.cs.ru.nl/B.Jacobs}

\thanks{The research leading to these results has received funding
  from the European Research Council under the European Union's
  Seventh Framework Programme (FP7/2007-2013) / ERC grant agreement
  nr.~320571}



\begin{document}

\maketitle

\begin{abstract}
This paper gives a systematic account of various metrics on
probability distributions (states) and on predicates. These metrics
are described in a uniform manner using the validity relation between
states and predicates. The standard adjunction between convex sets (of
states) and effect modules (of predicates) is restricted to convex
complete metric spaces and directed complete effect modules. This
adjunction is used in two state-and-effect triangles, for classical
(discrete) probability and for quantum probability.
\end{abstract}


\section{Introduction}\label{sec:intro}

Metric structures have a long history in program semantics, see the
overview book~\cite{deBakkerV96}. They occur naturally, for instance
on sequences, of inputs, outputs, or states. In complete metric spaces
solutions of recursive (suitably contractive) equations exist via
Banach's fixed point theorem. The Hausdorff distance on subsets is
used to model non-deterministic (possibilistic) computation. In
general, metrics can be used to measure to what extent computations
can be approximated, or are similar.

This paper looks at metrics on probability distributions (often called
states), as outcomes of probabilistic computations. Various such
metrics exist for measuring the (dis)similarity in behaviour between
computations,
see~\textit{e.g.}~\cite{BreugelW01,DesharnaisGJP04,BaldanBKK18}.  This
paper does not develop new applications, but contributes to the theory
behind distances in a probabilistic setting. In particular, it shows
how distances:
\begin{itemize}
\item arise in an abstract uniform way, both on
  distributions and on fuzzy predicates, see
  Equations~\eqref{eqn:statesvalidity}
  and~\eqref{eqn:predicatesvalidity} below for more information;

\item on distributions and on predicates can be related, via
  adjunctions in so-called state-and-effect triangles, see
  Diagrams~\eqref{diag:triangles:intro} below.
\end{itemize}

\noindent A salient feature of this paper is that it uniformly covers
standard distance functions, not only on \emph{classical} discrete
probability distributions, but also on \emph{quantum}
distributions. For discrete probability we use the familiar and
well-studied total variation distance, which is a special case of the
Kantorovich distance, see
\textit{e.g.}~\cite{GibbsSu02,BreugelHMW07,McIverMT15,MardarePP16}.
This total variation distance is investigated both on sets and on
metric spaces.  For quantum probability we use the well-known trace
distance for states (quantum distributions) on Hilbert spaces, and the
more general operator norm distance for states of von Neumann
algebras.  One contribution of this paper is a uniform description of
all these distances on states as `validity' distances.

In each of these cases we shall describe a validity relation $\models$
between states $\omega$ and predicates $p$, so that the validity
$\omega \models p$ is a number in the unit interval $[0,1]$. This
validity relation $\models$ plays a central role in the definition of
various distances. What we call the `validity' distance on states is
given by the supremum (join) over predicates $p$ in:
\begin{equation}
\label{eqn:statesvalidity}
\begin{array}{rcl}
d(\omega_{1}, \omega_{2})
& = &
{\displaystyle\bigvee}_{p}\, \big|\, \omega_{1}\models p - 
   \omega_{2}\models p \,\big|.
\end{array}
\end{equation}

\noindent In general, states are closed under convex combinations. We
shall thus study combinations of convex and complete metric spaces,
in a category $\ConvCMet$.

We also study metrics on predicates. The algebraic structure of
predicates will be described in terms of effect modules. Here we show
that suitably order complete effect modules are Archimedean, and thus
carry an induced metric, such that limits and joins of ascending
Cauchy sequences coincide. This result requires an extension of the
theory of effect modules, which is developed in
Section~\ref{sec:effects}.  In our main examples, we use fuzzy
predicates on sets and effects of von Neumann algebras as predicates;
their distance can also be formulated via validity $\models$, but now
using a join over states $\omega$ in:
\begin{equation}
\label{eqn:predicatesvalidity}
\begin{array}{rcl}
d(p_{1}, p_{2})
& = &
{\displaystyle\bigvee}_{\omega}\, \big|\, \omega\models p_{1} - 
   \omega\models p_{2} \,\big|.
\end{array}
\end{equation}

\noindent The `duality' between the distance
formulas~\eqref{eqn:statesvalidity} for states (distributions)
and~\eqref{eqn:predicatesvalidity} for predicates is a new insight.

A basic `dual' adjunction in a probabilistic setting is of the form
$\op{\EMod} \rightleftarrows \Conv$, between effect modules and convex
sets. Effect modules are the probabilistic analogues of Boolean
algebras, serving as `algebraic probabilistic logics' (see below for
details). Convex sets capture the algebraic structure of states. This
adjunction thus expresses the essentials of a probabilistic duality
between predicates and states. Since predicates are often called
`effects' in a quantum setting, one also speaks of a duality between
states and effects.

This paper restricts this adjunction to an adjunction
$\op{\DcEMod} \rightleftarrows \ConvCMet$ between \emph{directed
  complete} effect modules and convex \emph{complete metric}
spaces. This restriced adjunction is used in two `state-and-effect'
triangles, of the form:
\medskip
\begin{equation}
\label{diag:triangles:intro}
\vcenter{\xymatrix@C-1.8pc{
\op{\DcEMod}\ar@/^2ex/[rr] & \top & 
   \ConvCMet\ar@/^1.5ex/[ll] 
& \quad &
\op{\DcEMod}\ar@/^2ex/[rr] & \top & 
   \ConvCMet\ar@/^1.5ex/[ll] 
\\
& \Klf(\Dst)\ar[ul]^{\Pred}\ar[ur]_{\Stat} &
& &
& \op{\vNA}\ar[ul]^{\Pred}\ar[ur]_{\Stat} &
}}
\end{equation}

\noindent Details will be provided in Section~\ref{sec:triangles}.
Thus, the paper culminates in suitable order/metrically complete
versions of the state-and-effect triangles that emerge in the
effectus-theoretic~\cite{Jacobs15d,ChoJWW15b} description of state and
predicate transformer semantics for probability (see
also~\cite{Jacobs17a,Jacobs17b}).

\section{Distances between states}\label{sec:states}

This section will describe distance functions (metrics) on various
forms of probability distributions, which we collectively call
`states'. In separate subsections it will introduce discrete
probability distributions on sets and on metric spaces, and quantum
distributions on Hilbert spaces and on von Neumann algebras.  A
unifying formulation will be identified, namely what we call a
\emph{validity} formulation of the metrics involved, where the
distance between two states is expressed via a join over all
predicates using the validities of these predicates in the two states,
as in~\eqref{eqn:statesvalidity}.

\subsection{Discrete probability distributions on sets}\label{subsec:sets}

\mbox{}\\[-0.5em]

\noindent A finite discrete probability distribution on a set $X$ is
given by `probability mass' function $\omega \colon X \rightarrow
[0,1]$ with finite support and $\sum_{x}\omega(x) = 1$. This support
$\supp(\omega) \subseteq X$ is the set $\setin{x}{X}{\omega(x)\neq
  0}$. We sometimes simply say `distribution' instead of `finite
discrete probability distribution'. Often such a distribution is
called a `state'.  The `ket' notation $\ket{-}$ is useful to describe
specific distributions. For instance, on a set $X = \{a, b, c, d\}$ we
may write a distribution as $\omega = \frac{1}{2}\ket{a} +
\frac{1}{8}\ket{b} + \frac{3}{8}\ket{c}$. This corresponds to the
probability mass function $\omega \colon X \rightarrow [0,1]$ given by
$\omega(a) = \frac{1}{2}$, $\omega(b) = \frac{1}{8}$ and $\omega(c) =
\frac{3}{8}$.

We write $\Dst(X)$ for the set of distributions on a set $X$. The
mapping $X \mapsto \Dst(X)$ forms (part of) a well-known monad on the
category of sets, see
\textit{e.g.}~\cite{Jacobs11c,Jacobs16g,Jacobs17a} for additional
information, using the same notation as used here. We write
$\Kl(\Dst)$ for the associated Kleisli category, and $\EM(\Dst)$ for
the category of Eilenberg-Moore algebras. The latter may be identified
with \emph{convex sets}, that is, with sets in which formal convex
sums can be interpreted as actual sums. Thus we often write $\Conv =
\EM(\Dst)$; morphisms in $\Conv$ are `affine' functions, that preserve
convex sums. Convex sets have a rich history, going back
to~\cite{Stone49}, see~\cite[Remark~2.9]{KeimelP17} for an extensive
description.

\begin{definition}
\label{def:totvardist}
Let $\omega_{1}, \omega_{2} \in\Dst(X)$ be two distributions on the
same set $X$. Their \emph{total variation distance} $\tvd(\omega_{1},
\omega_{2})$ is the positive real number defined as:
\begin{equation}
\label{eqn:totvardist}
\begin{array}{rcl}
\tvd(\omega_{1}, \omega_{2})
& = &
\frac{1}{2}\displaystyle\sum_{x\in X} \big|\,\omega_{1}(x) - \omega_{2}(x)\,\big|.
\end{array}
\end{equation}
\end{definition}

The historical origin of this definition is not precisely clear.  It
is folklore that the total variation distance is a special case of the
`Kantorovich distance' (also known as `Wasserstein' or `earth mover's
distance') on distributions on metric spaces, when applied to discrete
metric spaces (sets), see Subsection~\ref{subsec:metric} below.

We leave it to the reader to verify that $\tvd$ is a metric on sets of
distributions $\Dst(X)$, and that its values are in the unit interval
$[0,1]$. 

\auxproof{
The distance $\tvd(\omega_{1}, \omega_{2})$ is in the unit interval
$[0,1]$ since:
$$\begin{array}{rcl}
\tvd(\omega_{1}, \omega_{2})
\hspace*{\arraycolsep}=\hspace*{\arraycolsep}
\frac{1}{2}\sum_{x} \big|\,\omega_{1}(x) - \omega_{2}(x)\,\big|
& \leq &
\frac{1}{2}\sum_{x} \big|\,\omega_{1}(x)\,\big| + \big|\,\omega_{2}(x)\,\big| 
\\
& \leq &
\frac{1}{2}\Big(\sum_{x} \omega_{1}(x) + \sum_{x}\omega_{2}(x)\Big) 
\\
& = &
\frac{1}{2}(1 + 1)
\hspace*{\arraycolsep}=\hspace*{\arraycolsep}
1.
\end{array}$$
}

\begin{example}
\label{ex:tvd}
Consider the sets $X = \{a,b\}$ and $Y=\{0,1\}$ with `joint'
distribution $\omega\in\Dst(X\times Y)$ given by $\omega =
\frac{1}{2}\ket{a,0} + \frac{1}{2}\ket{b,1}$. The first and second
marginal of $\omega$, written as $\omega_{1}\in\Dst(X)$ and
$\omega_{2} \in\Dst(Y)$, are: $\omega_{1} = \frac{1}{2}\ket{a} +
\frac{1}{2}\ket{b}$ and $\omega_{1} = \frac{1}{2}\ket{0} +
\frac{1}{2}\ket{1}$. We immediately see that $\omega$ is not the same
as the product $\omega_{1}\otimes\omega_{2}\in\Dst(X\times Y)$ of its
marginals, since $\omega_{1}\otimes\omega_{2} = \frac{1}{4}\ket{a,0} +
\frac{1}{4}\ket{a,1} + \frac{1}{4}\ket{b,0} + \frac{1}{4}\ket{b,1}$.
This means $\omega$ is `entwined', see~\cite{JacobsZ17,Jacobs17a}. One
way to associate a number with this entwinedness is to take the
distance between $\omega$ and the product of its marginals. It can be
computed as:
\[ \begin{array}{rcl}
\tvd\big(\omega, \,\omega_{1}\otimes\omega_{2}\big)
& = &
\frac{1}{2}\displaystyle\sum_{x\in X, y\in Y} 
   \big|\,\omega(x,y) - (\omega_{1}\otimes\omega_{2})(x,y)\,\big|
\\ 
& = &
\frac{1}{2}\displaystyle\textstyle\Big(|\frac{1}{2}-\frac{1}{4}| +
   |0-\frac{1}{4}| + |0-\frac{1}{4}| + |\frac{1}{2}-\frac{1}{4}|\Big)
\hspace*{\arraycolsep}=\hspace*{\arraycolsep}
\frac{1}{2}.
\end{array} \]
\end{example}

For a function $f \colon X \rightarrow \Dst(Y)$ there are two
associated `transformation' functions, namely state transformation
(aka.\ Kleisli extension) $f_{*}\colon \Dst(X) \rightarrow \Dst(Y)$
and predicate transformation $f^{*} \colon [0,1]^{Y} \rightarrow
[0,1]^{X}$. They are defined as:
\begin{equation}
\label{eqn:setstransformations}
\begin{array}{rclcrcl}
f_{*}(\omega)(y)
& = &
\sum_{x} f(x)(y)\cdot \omega(x)
& \qquad\mbox{and}\qquad &
f^{*}(q)(x)
& = &
\sum_{y} f(x)(y)\cdot q(y).
\end{array}
\end{equation}

\noindent Maps $p\in[0,1]^{X}$ are called (fuzzy) predicates on
$X$. In the special case where the outcomes $p(x)$ are in the
(discrete) subset $\{0,1\} \subseteq [0,1]$, the predicate $p$ is
called \emph{sharp}.  These sharp predicates correspond to subsets
$U\subseteq X$, via the indicator function $\indic{U} \colon X
\rightarrow \{0,1\}$.

For a state $\omega\in\Dst(X)$ we write $\omega\models p$ for the
\emph{validity} of predicate $p$ in state $\omega$, defined as the
expected value $\sum_{x}\omega(x) \cdot p(x)$ in $[0,1]$. Thus,
$\omega\models\indic{U}\, = \sum_{x\in U}\omega(x)$; the latter sum is
commonly written as $\omega(U)$. Further, the fundamental validity
transformation equality holds: $f_{*}(\omega) \models q \;=\; \omega
\models f^{*}(q)$.

We conclude this subsection with a standard redescription of the total
variation distance, see \textit{e.g.}~\cite{GibbsSu02,Villani09}. It
uses validity $\models$, as described above. Such `validity' based
distances will form an important theme in this paper. The proof of the
next result is standard but not trivial and is included in the
appendix, for the convenience of the reader.

\begin{proposition}
\label{prop:totvardist}
Let $X$ be an arbitrary set, with states $\omega_{1}, \omega_{2} \in
\Dst(X)$. Then:
\[ \begin{array}{rcccl}
\tvd\big(\omega_{1}, \omega_{2}\big)
& = &
\displaystyle\bigvee_{p\in[0,1]^{X}} 
   \Big|\,\omega_{1}\models p - \omega_{2}\models p\,\Big|
& = &
\displaystyle\max\limits_{U\subseteq X} \,
   \omega_{1}\models \indic{U} - \omega_{2}\models \indic{U}
\end{array}\eqno{\QEDbox} \]
\end{proposition}

\noindent We write maximum `$\max$' instead of join $\bigvee$ to
express that the supremum is actually reached by a subset (sharp
predicate). Completeness of the Kantorovich metric is an extensive
topic, but here we only need the following (standard) result. Since
there is a short proof, it is included.

\begin{lemma}
\label{lem:discretecompleteness}
If $X$ is a finite set, then $\Dst(X)$, with the total variation
distance $\tvd$, is a complete metric space.
\end{lemma}

\begin{myproof}
Let $X = \{x_{1}, \ldots, x_{N}\}$ and $\omega_{i}\in\Dst(X)$ be a
Cauchy sequence. For each $n$ we have $\big|\omega_{i}(x_{n}) -
\omega_{j}(x_{n})\big| \leq 2\cdot\tvd(\omega_{i},
\omega_{j})$. Hence, the sequence $\omega_{i}(x_{n}) \in [0,1]$ is
Cauchy too, say with limit $r_{n}$. Take $\omega = \sum_{n}
r_{n}\ket{x_{n}} \in \Dst(X)$. This is the limit of the
$\omega_{i}$. \QED
\end{myproof}

\subsection{Discrete probability distributions on metric 
   spaces}\label{subsec:metric}

\mbox{}\\[-0.5em]

\noindent A metric $d$ on a set $X$ is called 1-bounded if it takes
values in the unit interval $[0,1]$, that is, if it has type $d\colon
X\times X \rightarrow [0,1]$. We write $\Met$ for the category with
such 1-bounded metric spaces as objects, and with non-expansive
functions $f$ between them, satisfying $d(f(x), f(y)) \leq
d(x,y)$. From now on we assume that all metric spaces in this paper
are 1-bounded. For example, each set carries a discrete metric, where
points $x,y$ have distance $0$ if they are equal, and $1$ otherwise.


For a metric space $X$ and two functions $f,g \colon A \rightarrow X$
from some set $A$ to $X$ there is the \emph{supremum} distance given
by:
\begin{equation}
\label{eqn:supmet}
\begin{array}{rcl}
\spd(f,g)
& = &
\displaystyle \bigvee_{a\in A} d\big(f(a), g(a)\big).
\end{array}
\end{equation}

A `metric predicate' on a metric space $X$ is a non-expansive function
$p\colon X \rightarrow [0,1]$. These predicates carry the above
supremum distance $\spd$. We use them in the following definition of
Kantorovich distance, which transfers the validity description of
Proposition~\ref{prop:totvardist} to the metric setting.

\begin{definition}
\label{def:kantorovich}
Let $\omega_{1},\omega_{2}$ be two discrete distributions on (the
underlying set of) a metric space $X$. The \emph{Kantorovich}
distance between them is defined as:
\begin{equation}
\label{eqn:kantorovich}
\begin{array}{rcl}
\kvd(\omega_{1}, \omega_{2})
& = &
\displaystyle\bigvee_{p\in\Met(X,[0,1])} 
   \Big|\,\omega_{1}\models p - \omega_{2}\models p\,\Big|.
\end{array}
\end{equation}

\noindent This makes $\Dst(X)$ a (1-bounded) metric space.
\end{definition}

The Kantorovich-Wasserstein duality Theorem gives an equivalent
description of this distance in terms of joint states and `couplings',
see~\cite{Lindvall92,Villani09} for details. Here we concentrate on
relating the Kantorovich distance to the monad structure of
distributions. The next lemma collects some basic, folkore facts.

\begin{lemma}
\label{lem:kantorovich}
Let $X,Y$ be metric spaces.
\begin{enumerate}
\item \label{lem:kantorovichUnit} The unit function $\eta \colon X
  \rightarrow \Dst(X)$ given by $\eta(x) = 1\ket{x}$ is non-expansive.

\item \label{lem:kantorovichMap} For each non-expansive function $f\colon X
  \rightarrow \Dst(Y)$ the corresponding state transformer $f_{*}
  \colon \Dst(X) \rightarrow \Dst(Y)$
  from~\eqref{eqn:setstransformations} is non-expansive. 

As special cases, the multiplication map $\mu = (\idmap)_{*} \colon
\Dst(\Dst(X)) \rightarrow \Dst(X)$ of the monad $\Dst$ is
		non-expansive, and validity $((-) \models p) = p_{*} \colon \Dst(X)
\rightarrow \Dst(2) = [0,1]$ in its first argument as well.

\item \label{lem:kantorovichSubst} If $f\colon X \rightarrow \Dst(Y)$ and
  $q\colon Y \rightarrow [0,1]$ are non-expansive, then so is
  $f^{*}(q) \colon X \rightarrow [0,1]$. Moreover, the function $f^{*}
  \colon \Met(Y, [0,1]) \rightarrow \Met(X, [0,1])$ is itself
  non-expansive, wrt.\ the supremum distance~\eqref{eqn:supmet}. 

		As a result, validity $(\omega \models (-)) = \omega^{*} \colon
\Met(X,[0,1]) \rightarrow \Met(1,[0,1]) = [0,1]$ is non-expansive in
its second argument too.

\item \label{lem:kantorovichSubstConvex} Taking convex combinations of
  distributions $\sigma_{i}, \tau_{i}$ satisfies: for $r+s = 1$,
$$\begin{array}{rcl}
\kvd\big(r\cdot \sigma_{1} + s\cdot\sigma_{2}, \;
   r\cdot \tau_{1} + s\cdot\tau_{2}\big)
& \leq &
r\cdot \kvd(\sigma_{1}, \tau_{1}) + s\cdot \kvd(\sigma_{2}, \tau_{2}).
\end{array}$$

\end{enumerate}
\end{lemma}

\begin{myproof}
We do points~\eqref{lem:kantorovichUnit}
and~\eqref{lem:kantorovichSubstConvex} and leave the others to the
reader. The crucial fact that we use 
for~\eqref{lem:kantorovichUnit} is that the unit map $\eta \colon X
	\rightarrow \Dst(X)$ is non-expansive: $(\eta(x) \models p) \,=\,
p(x)$. Hence we are done because the join in~\eqref{eqn:kantorovich}
is over non-expansive functions $p$ in:
$$\begin{array}{rcl}
\kvd(\eta(x_{1}), \eta(x_{2}))
\hspace*{\arraycolsep}=\hspace*{\arraycolsep}
{\displaystyle\bigvee}_{p}\,
   \Big|\,\eta(x_{1})\models p - \eta(x_{2})\models p\,\Big|
& = &
{\displaystyle\bigvee}_{p}\, \Big|\,p(x_{1}) - p(x_{2})\,\Big|
\\
& \leq &
{\displaystyle\bigvee}_{p}\, d(x_{1},x_{2})
\\
& = &
d(x_{1},x_{2}).
\end{array}$$

\auxproof{
For a Kleisli map $f\colon X \rightarrow \Dst(Y)$ we have:
$$\begin{array}{rcl}
\kvd\big(f_{*}(\omega_{1}), f_{*}(\omega_{2})\big)
& = &
{\displaystyle\bigvee}_{q\in\Pred(Y)} 
   \Big|\,f_{*}(\omega_{1})\models q - f_{*}(\omega_{2})\models q\,\Big|
\\
& = &
{\displaystyle\bigvee}_{q\in\Pred(Y)} 
   \Big|\,\omega_{1}\models f^{*}(q) - \omega_{2}\models f^{*}(q)\,\Big|
\\
& \leq &
{\displaystyle\bigvee}_{p\in\Pred(X)} 
   \Big|\,\omega_{1}\models p - \omega_{2}\models p\,\Big|
\hspace*{\arraycolsep}=\hspace*{\arraycolsep}
\kvd(\omega_{1}, \omega_{2}).
\end{array}$$

An explicit proof that the multiplication map $\mu$ is non-expansive:
$$\begin{array}{rcl}
\kvd\big(\mu(\Omega), \mu(\Omega')\big)
& = &
\displaystyle\bigvee_{p\in\Pred(X)} 
   \Big|\,\mu(\Omega)\models p - \mu(\Omega')\models p\,\Big|
\\
& = &
\displaystyle\bigvee_{p\in\Pred(X)} 
   \Big|\,\Omega \models \big((-)\models p\big) - 
      \Omega' \models \big((-)\models p\big)\,\Big|
\\
& \leq &
\displaystyle\bigvee_{Q\in\Pred(\Dst(X))} 
   \Big|\,\Omega\models Q - \Omega'\models Q\,\Big|
\\
& = &
\kvd(\Omega, \Omega').
\end{array}$$

We see that $f^{*}(q)$ is non-expansive via the
transformations validity equation:
$$\begin{array}{rcl}
\big|\,f^{*}(q)(x_{1}) - f^{*}(q)(x_{2})\,\big|
& = &
\big|\, \sum_{y} f(x_{1})(y)\cdot q(y) - \sum_{y} f(x_{2})(y)\cdot q(y) \,\big|
\\
& = &
\big|\, f(x_{1}) \models q - f(x_{2}) \models q \,\big|
\\
& \leq &
{\displaystyle\bigvee}_{p}\, 
   \big|\, f(x_{1}) \models p - f(x_{2}) \models p \,\big|
\\
& = &
\kvd\big(f(x_{1}), f(x_{2})\big)
\\
& \leq &
d_{X}(x_{1}, x_{2}) \qquad \mbox{since $f$ is non-expansive.}
\end{array}$$

\noindent It is not hard to see that the predicate transformer $f^{*}$
is non-expansive:
$$\begin{array}{rcl}
\spd\big(f^{*}(q_{1}), f^{*}(q_{2})\big)
& = &
{\displaystyle\bigvee}_{x}\, \big|\, f^{*}(q_{1})(x) - f^{*}(q_{2})(x) \,\big|
\\
& = &
{\displaystyle\bigvee}_{x}\, 
   \big|\, \sum_{y} f(x)(y)\cdot q_{1}(y) - \sum_{y} f(x)(y)\cdot q_{2}(y) \,\big|
\\
& \leq &
{\displaystyle\bigvee}_{x}\, \sum_{y} f(x)(y)\cdot 
   \big|\, q_{1}(y) - q_{2}(y) \,\big|
\\
& \leq &
{\displaystyle\bigvee}_{x}\, \sum_{y} f(x)(y)\cdot {\displaystyle\bigvee}_{z}\,
   \big|\, q_{1}(z) - q_{2}(z) \,\big|
\\
& = &
{\displaystyle\bigvee}_{x}\, \sum_{y} f(x)(y)\cdot \spd(q_{1}, q_{2})
\\
& = &
{\displaystyle\bigvee}_{x}\, \spd(q_{1}, q_{2})
\\
& = &
\spd(q_{1}, q_{2}).
\end{array}$$
}

\noindent For point~\eqref{lem:kantorovichSubstConvex} we first notice
that for $\Omega\in\Dst^{2}(X)$ and $p\colon X \rightarrow [0,1]$,
$$\begin{array}{rcl}
	(\mu(\Omega) \models p)
\hspace*{\arraycolsep}\,=\,\hspace*{\arraycolsep}
\sum_{x} \mu(\Omega)(x) \cdot p(x)
& = &
\sum_{x} \big(\sum_{\omega} \Omega(\omega)\cdot \omega(x)\big) \cdot p(x)
\\
& = &
\sum_{\omega} \Omega(\omega)\cdot \big(\sum_{x}\omega(x) \cdot p(x)\big)
\\
& = &
\sum_{\omega} \Omega(\omega)\cdot \big(\omega\models p\big)
\\
& = &
	(\Omega \models \big((-)\models p\big)),
\end{array}$$

\noindent where $\mu$ is the multiplication map
defined in Lemma~\ref{lem:kantorovich}\eqref{lem:kantorovichMap}, 
and $((-)\models p) \colon \Dst(X) \rightarrow [0,1]$ is
used as (non-expansive) predicate on $\Dst(X)$. Hence for $r,s\in
[0,1]$ with $r+s = 1$,
$$\begin{array}[b]{rcl}
\lefteqn{\kvd\big(r\cdot \sigma_{1} + s\cdot\sigma_{2}, \;
   r\cdot \tau_{1} + s\cdot\tau_{2}\big)}
\\
& = &
\kvd\big(\mu(r\ket{\sigma_{1}} + s\ket{\sigma_{2}}), \;
   \mu(r\ket{\tau_{1}} + s\ket{\tau_{2}})\big)
\\
& = &
{\displaystyle\bigvee}_{p}\, \big|\,
   \mu(r\ket{\sigma_{1}} + s\ket{\sigma_{2}}) \models p -
   \mu(r\ket{\tau_{1}} + s\ket{\tau_{2}}) \models p \,\big|
\\
& = &
{\displaystyle\bigvee}_{p}\, \big|\,
   r\ket{\sigma_{1}} + s\ket{\sigma_{2}} \models \big((-)\models p\big) -
   r\ket{\tau_{1}} + s\ket{\tau_{2}} \models \big((-)\models p\big) \,\big|
\\
& = &
{\displaystyle\bigvee}_{p}\, \big|\,
   r\cdot (\sigma_{1}\models p) + s\cdot(\sigma_{2}\models p) -
   r\cdot (\tau_{1}\models p) + s\cdot(\tau_{2}\models p) \,\big|
\\
& \leq &
{\displaystyle\bigvee}_{p}\, r\cdot \big|\,
   \sigma_{1}\models p - \tau_{1}\models p \,\big| + 
   {\displaystyle\bigvee}_{p}\, s\cdot \big|\,
      \sigma_{2}\models p - \tau_{2}\models p \,\big|
\\
& = &
r\cdot \kvd(\sigma_{1}, \tau_{1}) + s\cdot \kvd(\sigma_{2}, \tau_{2}).
\end{array}\eqno{\QEDbox}$$
\end{myproof}

\begin{corollary}
\label{cor:Dstlift}
The monad $\Dst$ on $\Sets$ lifts to a monad, also written as $\Dst$,
on the category $\Met$, and commutes with forgetful functors, as in:
\begin{equation}
\label{diag:distlift}
\vcenter{\xymatrix@R-0.5pc{
\Met\ar[d]\ar[rr]^-{\Dst} & & \Met\ar[d]
\\
\Sets\ar[rr]^-{\Dst} & & \Sets
}}
\end{equation}

\noindent We write $\ConvMet$ for the category $\EM(\Dst)$ of
Eilenberg-Moore algebras of this lifted monad, with `convex metric
spaces' as objects, see below.
\end{corollary}

The lifting~\eqref{diag:distlift} can be seen as a finite version of a
similar lifting result for the `Kantorovich' functor $\mathcal{K}$
in~\cite{BreugelHMW07}.  This $\mathcal{K}(X)$ captures the tight
Borel probability measures on a metric space $X$.  The above
lifting~\eqref{diag:distlift} is a special case of the generic lifting
of functors on sets to functors on metric spaces described
in~\cite{BaldanBKK14} (see esp.\ Example~3.3).

The category $\ConvMet = \EM(\Dst)$ of the monad
$\Dst\colon\Met\rightarrow\Met$ contains \emph{convex metric spaces},
consisting of:
\begin{enumerate}
\item a convex set $X$, that is, a set $X$ with an Eilenberg-Moore
  algebra $\alpha\colon\Dst(X) \rightarrow X$ of the distribution
  monad $\Dst$ on $\Sets$;

\item a metric $d_{X}\colon X\times X \rightarrow [0,1]$ on $X$;

\item a connection between the convex and the metric structure, via
  the requirement that the algebra map $\alpha\colon \Dst(X)
  \rightarrow X$ is non-expansive: $d_{X}(\alpha(\omega_{1}),
  \alpha(\omega_{2})) \leq \kvd(\omega_{1}, \omega_{2})$, for all
  distributions $\omega_{1}, \omega_{2}\in\Dst(X)$.
\end{enumerate}

\noindent The maps in $\ConvMet$ are both affine and non-expansive.
We shall write $\ConvCMet \hookrightarrow \ConvMet$ for the full
subcategory of convex \emph{complete} metric spaces.

\begin{example}
\label{ex:unitconvmet}
The unit interval $[0,1]$ is a convex metric space, via its standard
(Euclidean) metric, and its standard convex structure, given by the
algebra map $\alpha\colon\Dst([0,1]) \rightarrow [0,1]$ defined by
the `expected value' operation:
\[ \begin{array}{rclcrcl}
\alpha(\omega)
& = &
\sum_{x\in\R} \omega(x)\cdot x
& \qquad\mbox{that is}\qquad &
\alpha\big(\sum_{i}r_{i}\ket{x_i}\big)
& = &
\sum_{i} r_{i}\cdot x_{i}.
\end{array} \]

\noindent The identity map $\idmap \colon [0,1] \rightarrow [0,1]$ is
a predicate on $[0,1]$ that satisfies:
\[ \begin{array}{rcccccl}
	(\omega\models \idmap)
& = &
\sum_{x} \omega(x) \cdot \idmap(x)
& = &
\sum_{x} \omega(x) \cdot x
& = &
\alpha(\omega).
\end{array} \]

\noindent This allows us to show that $\alpha$ is non-expansive:
\[ \begin{array}{rcl}
\big|\,\alpha(\omega_{1}) - \alpha(\omega_{2})\,\big|
& = &
\big|\,\omega_{1}\models \idmap - \omega_{2} \models \idmap\,\big| \\
& \leq &
{\displaystyle\bigvee}_{\!p}\; \big|\,\omega_{1}\models p - \omega_{2} \models p\,\big|
\hspace*{\arraycolsep}=\hspace*{\arraycolsep}
\kvd(\omega_{1}, \omega_{2}).
\end{array} \]

\noindent In fact, we can see this as a special case of
non-expansiveness of multiplication maps $\mu$ from
Lemma~\ref{lem:kantorovich}~\eqref{lem:kantorovichMap}: indeed,
$\Dst(2) \cong [0,1]$, for the two-element set $2 = \{0,1\}$, and the
algebra $\alpha\colon\Dst([0,1]) \rightarrow [0,1]$ corresponds to the
multiplication $\mu\colon\Dst(\Dst(2)) \rightarrow \Dst(2)$.
\end{example}

\subsection{Density matrices on Hilbert spaces}\label{subsec:density}

\mbox{}\\[-0.5em]

\noindent The analogue of a probability distribution in quantum theory
is often simply called a state. We first consider states of Hilbert
spaces (over~$\C$), and consider the more general (and abstract) situation of
states of von Neumann algebras in subsection~\ref{subsec:vNA}.

A state of a Hilbert space $\mathscr{H}$ is a density operator, that
is, it is a positive linear map $\varrho \colon \mathscr{H}
\rightarrow \mathscr{H}$ whose trace is one: $\tr(\varrho) = 1$.
Recall that the trace of a positive operator $T\colon
\mathscr{H}\to\mathscr{H}$ is given by $\tr(T)=\sum\tuple{e_i,
  T(e_i)}$, where~$(e_i)_i$ is any orthonormal basis
for~$\mathscr{H}$; this value $\tr(T)$ does not depend on the choice
of basis~$(e_i)_i$, but might
equal~$+\infty$~\cite[Def.~2.51]{Alfsen2012}. The same formula also
works for when~$T$ is not necessarily positive, but bounded with
$\tr(\left|T\right|)<\infty$ --- where $\left|T\right| \coloneqq
\sqrt{T^\dagger T}$ and $T^\dagger$ is the adjoint of~$T$ and where
the square root is determined as the unique positive operator $B$ with
$BB = T^\dagger T$.  Such~$T$, which are aptly called
\emph{trace-class operators}, always have finite trace:
$\tr(T)<\infty$, see~\cite[Def.~2.5\{4,6\}]{Alfsen2012}.
When~$\mathscr{H}$ is finite dimensional, any operator~$T\colon
\mathscr{H}\to\mathscr{H}$ is trace-class, and when represented as a
matrix, its trace can be computed as the sum of all elements on the
diagonal.  If~$T$ is a density operator, then the associated matrix is
called a density matrix.  We refer for more information to for
instance~\cite{Alfsen2012}, and
to~\cite{NielsenC00,RieffelP11,YanofskyM08} for the finite-dimensional
case.

A linear map $A\colon \mathscr{H} \rightarrow \mathscr{H}$ is called
\emph{self-adjoint} if $A = A^{\dagger}$ and \emph{positive} if it is
of the form $A = BB^{\dagger}$. This yields a partial order, with $A
\leq B$ iff $B-A$ is positive. A predicate on $\mathscr{H}$ is a
linear map $p\colon \mathscr{H} \rightarrow \mathscr{H}$ with $0 \leq
p \leq \idmap$. It is called \emph{sharp} (or a \emph{projection}) if
$p^{2} = p$. Predicates are also called \emph{effects}. We write
$\Ef(\mathscr{H})$ for the set of effects of $\mathscr{H}$.  For a
state $\varrho$ of $\mathscr{H}$ the validity $\varrho\models p$ is
defined as the trace $\tr(\varrho\,p)$.  To make sense of this
definition we should mention that the product $AB$ of bounded
operators $A,B\colon \mathscr{H}\to\mathscr{H}$ is trace-class when
either~$A$ or~$B$ is trace-class~\cite[Def.~2.54]{Alfsen2012} ---
so~$\varrho\, p$ is trace-class because~$\varrho$ is.

\begin{definition}
\label{def:tracedist}
Let $\varrho_{1}, \varrho_{2}$ be two quantum states of the same Hilbert
space. The trace distance $\trd(\varrho_{1}, \varrho_{2})$ between them is
defined as:
\begin{equation}
\label{eqn:tracedist}
\begin{array}{rcccl}
\trd(\varrho_{1}, \varrho_{2})
& = &
\frac{1}{2}\tr\big(\big|\,\varrho_{1} - \varrho_{2}\,\big|\big)
& = &
\frac{1}{2}\tr\big(\sqrt{(\varrho_{1}-\varrho_{2})^{\dag}(\varrho_{1}-\varrho_{2})}\big).
\end{array}
\end{equation}
\end{definition}

This definition involves the square root of a positive operator~$B$.
With the examples below in mind it is worth pointing out that in the
finite-dimensional case --- when~$B$ is essentially a positive matrix
--- the square root of~$B$ can be computed by first diagonalising the
matrix $B = VDV^{\dag}$, where $D$ is a diagonal matrix; then one
forms the diagonal matrix $\sqrt{D}$ by taking the square roots of the
elements on the diagonal in $D$; finally the square root of $B$ is
$V\sqrt{D}V^{\dag}$.

The trace distance $\trd$ is an extension of the total variation
distance $\tvd$: given two discrete distributions $\omega_{1},
\omega_{2}$ on the same set, then the union of their supports
$\supp(\omega_{1}) \cup \supp(\omega_{2})$ is a finite set, say with
$n$ elements. We can represent $\omega_{1}, \omega_{2}$ via diagonal
$n\times n$ matrices as density operators $\widehat{\omega_1},
\widehat{\omega_2}$. They are states, by construction. Then
$\trd(\widehat{\omega_1}, \widehat{\omega_2}) = \tvd(\omega_{1},
\omega_{2})$.

\begin{example}
\label{ex:trd}
We describe the quantum analogue of Example~\ref{ex:tvd}, involving
the `Bell' state. As a vector in $\C^{2}\otimes\C^{2}$ the Bell state
is usually described as $\ket{b} = \frac{1}{\sqrt{2}}(\ket{00} +
\ket{11})$. The corresponding density matrix
$\beta = \ket{b}\bra{b}$ is the following $4\times 4$ matrix.
\[ \begin{array}{rcccl}
\beta
& = &
\frac{1}{\sqrt{2}}
\left(\begin{smallmatrix}
1 \\ 0 \\ 0 \\ 1
\end{smallmatrix}\right)
\frac{1}{\sqrt{2}}
\left(\begin{smallmatrix}
1 & 0 & 0 & 1
\end{smallmatrix}\right)
& = &
\frac{1}{2}\left(\begin{smallmatrix}
1 & 0 & 0 & 1 \\
0 & 0 & 0 & 0 \\
0 & 0 & 0 & 0 \\
1 & 0 & 0 & 1 \\
\end{smallmatrix}\right)
\end{array} \]

\noindent Its two marginals
    $\beta_{1}, \beta_{2}$
    (in this context usually called \emph{reduced density operators}
    and obtained by taking \emph{partial traces},
    see~\S2.4.3 of~\cite{NielsenC00}) 
are equal $2\times 2$ matrices, namely:
\[ \begin{array}{rccclcrcl}
\beta_{1}
& = &
\beta_{2}
& = &
\frac{1}{2}\left(\begin{smallmatrix}
1 & 0 \\
0 & 1 \\
\end{smallmatrix}\right)
& \qquad\mbox{so that}\qquad &
\beta_{1}\otimes\beta_{2}
& = &
\frac{1}{4}\left(\begin{smallmatrix}
1 & 0 & 0 & 0 \\
0 & 1 & 0 & 0 \\
0 & 0 & 1 & 0 \\
0 & 0 & 0 & 1 \\
\end{smallmatrix}\right)
\end{array} \]

\noindent The product state $\beta_{1}\otimes\beta_{2}$ is obtained as
Kronecker product, see \textit{e.g.}~\cite{NielsenC00}.

We can now ask the same question as in Example~\ref{ex:tvd}, namely
what is the distance between the Bell state $\beta$ and the product of
its marginals. We recall that the Bell state is `maximally entangled'
and that the quantum theory allows, informally stated, higher levels
of entanglement than in classical probability theory. Hence we expect
an outcome that is higher than the value $\frac{1}{2}$ obtained in
Example~\ref{ex:tvd} for the classical maximally entwined state.

The key steps are:
\[ \begin{array}{rclcrcl}
\beta - \beta_{1}\otimes\beta_{2}
& = &
\left(\begin{smallmatrix}
\nicefrac{1}{4} & 0 & 0 & \nicefrac{1}{2} \\
0 & -\nicefrac{1}{4} & 0 & 0 \\
0 & 0 & -\nicefrac{1}{4} & 0 \\
\nicefrac{1}{2} & 0 & 0 & \nicefrac{1}{4} \\
\end{smallmatrix}\right)
& \qquad\mbox{so that}\qquad &
\big|\beta - \beta_{1}\otimes\beta_{2}\big|
& = &
\left(\begin{smallmatrix}
\nicefrac{1}{2} & 0 & 0 & \nicefrac{1}{2} \\
0 & \nicefrac{1}{4} & 0 & 0 \\
0 & 0 & \nicefrac{1}{4} & 0 \\
\nicefrac{1}{2} & 0 & 0 & \nicefrac{1}{2} \\
\end{smallmatrix}\right)
\end{array} \]

\noindent Hence:
\[ \begin{array}{rcccccl}
\trd\big(\beta, \,\beta_{1}\otimes\beta_{2}\big)
& = &
\frac{1}{2}\tr\big(\big|\beta - \beta_{1}\otimes\beta_{2}\big|\big)
& = &
\frac{1}{2}\Big(\nicefrac{1}{2} + \nicefrac{1}{4} + 
   \nicefrac{1}{4} + \nicefrac{1}{2}\Big)
& = &
\frac{3}{4}.
\end{array} \]

\noindent In the earlier version of this paper~\cite{Jacobs17c} these
distance computations are generalised to $n$-ary products, both for
classical and for quantum states. Both distances then tend to $1$, as
$n$ goes to infinity, but the classical distance is one step behind,
via formulas $\frac{2^{n-1}-1}{2^{n-1}}$ versus
$\frac{2^{n}-1}{2^{n}}$. Here we only consider $n=2$.
\end{example}

The following result is a quantum analogue of
Proposition~\ref{prop:totvardist}. Our formulation generalises the
standard formulation of \eg~\cite[\S9.2]{NielsenC00} and its proof to
arbitrary, not necessarily finite-dimensional Hilbert spaces.  We will
see an even more general version involving von Neumann algebras later
on.

\begin{proposition}
\label{prop:trace}
For states $\varrho_{1}, \varrho_{2}$ on the same Hilbert space $\mathscr{H}$,
\[ \begin{array}[b]{rcccl}
\trd(\varrho_{1}, \varrho_{2})
& = &
\displaystyle\bigvee_{p\in \Ef(\mathscr{H})} \big|\, \varrho_{1} \models p - 
   \varrho_{2} \models p \,\big|
& = &
\max\limits_{s\in \Ef(\mathscr{H})\text{ sharp }} \,(\varrho_{1} \models s) - 
   (\varrho_{2} \models s) .
\end{array}\eqno{\QEDbox} \]
\end{proposition}

\noindent As before, the maximum means the supremum is actually
reached by a sharp effect. The proof of this result is in the
appendix.

\subsection{Preliminaries on von Neumann algebras}

\mbox{}\\[-0.5em]

\noindent
Our final example of a distance function
requires a short introduction to von Neumann algebras.
We do not however pretend to explain the basics of the theory of von
Neumann algebras here; for this we refer
to~\cite{Kadison2015} (and~\cite{bram}). We just recall some elementary 
definitions and facts which are relevant here. 
\newcommand{\scrA}[0]{\mathscr{A}}

To define von Neumann algebras
we must speak about unital $C^*$-algebras first.
\begin{definition}
A \emph{unital $C^*$-algebra}~$\scrA$
is a complex vector space endowed with:
\begin{enumerate}
\item 
an associative  multiplication
that is linear in both coordinates;
\item
an element $1$, called unit, 
such that $1\cdot a=a=a\cdot 1$ for all~$a\in \scrA$;
\item
a unary operation $(\,\cdot\,)^*$,
called involution,
such that $(a^*)^*=a$,
		$(ab)^*=b^*a^*$,
		$(\lambda a)^*=\overline{\lambda}a^*$,
		and $(a+b)^*=a^*+b^*$
		for all~$a,b\in\scrA$ and~$\lambda\in\C$;
\item
a complete norm, $\left\|\,\cdot\,\right\|$,
with $\left\|ab\right\|\leq \left\|a\right\|\left\|b\right\|$
and $\left\|a^*a\right\|=\left\|a\right\|^2$
for all~$a,b\in\scrA$.
\end{enumerate}
\end{definition}
Two types of elements deserve special mention: an element~$a$ of a
unital $C^*$-algebra~$\scrA$ is called \emph{self-adjoint} when~$a^*=a$, and
\emph{positive} when~$a\equiv b^*b$ for some~$b\in
\scrA$.\footnote{In~\cite{Kadison2015} a different but in the end
  equivalent definition of ``positive'' is used, see Theorem~4.2.6
  of~\cite{Kadison2015}.}  Elementary matters relating to self-adjoint
elements are usually easily established: the reader should have no
trouble verifying, for example, that every element~$a$ of a
unital $C^*$-algebra~$\scrA$ can be written as $a\equiv b+ic$ for unique
self-adjoint~$b,c\in\scrA$ (namely, $b=\frac{1}{2}(a+a^*)$ and
$c=\frac{1}{2i}(a-a^*)$.)  On the other hand, the everyday properties
of the positive elements are often remarkably difficult to prove from
basic principles, such as the facts that the sum of positive elements
is positive, that the set~$\scrA_+$ of positive elements of~$\scrA$ is
norm closed (see parts (iii) and~(i) of Theorem~4.2.2
of~\cite{Kadison2015}), that every positive element~$a\in\scrA_+$ has
a unique positive square root, $\sqrt{a}$ (see Theorem~4.2.6(ii)
of~\cite{Kadison2015}), and that every self-adjoint element~$a$
of~$\scrA$ may be written uniquely as $a\equiv b-c$
where~$b,c\in\scrA_+$ with~$bc=0$ (see Proposition~4.2.3(iii)
of~\cite{Kadison2015}).

The elements of a unital $C^*$-algebra
are ordered by $a\leq b$ when $b-a$ is positive.
 We write
$[0,1]_{\mathscr{A}} \subseteq \mathscr{A}$ for subset of effects $0
\leq e \leq 1$; they will be used as quantum predicates. Such an
effect $e$ is called sharp (or a projection) if $e^{2} = e$.

\begin{definition}
A unital $C^{*}$-algebra $\mathscr{A}$ is a \emph{von Neumann} algebra
(aka.\ $W^*$-algebra) if firstly the unit interval
$[0,1]_{\mathscr{A}}$ is a directed complete partial order (dcpo), and
secondly the positive linear functionals $\omega\colon
    \mathscr{A}\to\C$ with $\omega(1)=1$
    that preserve these (directed) suprema separate the
elements of~$[0,1]_{\mathscr{A}}$.
    This means that $e_1,e_2\in[0,1]_\mathscr{A}$
are equal provided
    that $\omega(e_1)=\omega(e_2)$
    for all such~$\omega$.
\end{definition}
There are several
equivalent alternative definitions of the notion of `von Neumann
algebra', but this one, essentially due to Kadison
(see~\cite{Kadison56}), is most convenient here.

For the purposes of this paper
we consider as morphisms $f\colon \mathscr{A} \rightarrow \mathscr{B}$
between von Neumann algebras: linear maps which are unital (that is,
$f(1) = 1$), positive ($a \geq 0$ implies $f(a) \geq 0$) and
normal.\footnote{It is not difficult to see that a 
positive morphism between von Neumann algebras sends self-adjoint elements
to self-adjoint elements, and preserves the involution.}
The latter normality requirement means that the restriction $f
\colon [0,1]_{\mathscr{A}} \rightarrow [0,1]_{\mathscr{B}}$ preserves
directed joins (\ie~is Scott-continuous). This yields a category
$\vNA$ of von Neumann algebras. It occurs naturally in opposite form,
as $\op{\vNA}$.\footnote{Depending on the context other choices
of morphisms between von Neumann algebras may be more appropriate.
The proper `structure preserving maps', for example,  preserve 
multiplication too, and form a strict subcategory
$\vNA'$ of~$\vNA$, similar to how
$\Sets$ forms a strict subcategory of~$\Kl(\Dst)$.
Further, if we had wished to work with tensor products
of von Neumann algebras,
we had required the morphisms in~$\vNA$
to be not just positive but \emph{completely positive},
see~\cite{Stinespring1955}.}

Each non-zero\footnote{The unique morphism $\mathscr{A}\to\{0\}$ 
is unital, because it sends~$1$ to~$1=0$,
but has operator norm~$0$.} morphism $f$ in $\vNA$ has operator norm equal to 1,
\ie~$\opnorm{f} = 1$, where $\opnorm{f} =
\bigvee\set{\|f(x)\|}{\|x\|=1}$. Below we apply the operator norm to a
(pointwise) difference $\opnorm{f-g}$ of parallel morphisms $f,g$ in
$\vNA$.  Using $\opnorm{f-g}$ as distance, each homset of $\vNA$ is a
complete metric space.\footnote{%
Here is a proof that~$\vNA(\mathscr{A},\mathscr{B})$
is complete:
We must show that a Cauchy sequence
$f_1,f_2,\dotsc$
in $\vNA(\mathscr{A},\mathscr{B})$
converges.
By Theorem~1.5.6 of~\cite{Kadison2015},
the sequence $f_1,f_2,\dotsc$
$\opnorm{-}$-converges
to a bounded linear map $f\colon \mathscr{A}\to\mathscr{B}$.
It is clear that~$f$ will be unital,
and positive (since
the norm-limit of positive elements of~$\mathscr{B}$ is again positive,
see Theorem~4.2.2 of~\cite{Kadison2015}),
so it remains to be shown that~$f$ is normal.
Given directed~$D\subseteq[0,1]_{\mathscr{A}}$
we must show that $\bigvee_{d\in D} f(d)
= f(\bigvee D)$.
For this
it suffices to show that
$\bigvee_{d\in D} \omega(f(d))\equiv
\omega(\bigvee_{d\in D} f(d))=\omega(f(\bigvee D))$
for all positive normal linear functionals $\omega\colon \mathscr{B}\to\C$,
which is the case when~$\omega\circ f$ is normal.
But since~$\omega\circ f_1,\,\omega\circ f_2,\,\cdots$
$\opnorm{-}$-converges
to $\omega \circ f$,
this is indeed so
(because the predual of~$\mathscr{B}$ is complete, 
see the text under Definition~7.4.1
of~\cite{Kadison2015}.)}

\subsection{States of von Neumann algebras}\label{subsec:vNA}

\mbox{}\\[-0.5em]

A state of a von Neumann algebra $\mathscr{A}$ is a morphism
$\varrho\colon\mathscr{A} \rightarrow \C$ in $\vNA$. We write
$\Stat(\mathscr{A}) = \Hom(\mathscr{A},\C)$ for the set of states; it
is easy to see that it is a convex set. For an effect $e\in
[0,1]_{\mathscr{A}}$ we write $\varrho\models e$ for the value
$\varrho(e) \in [0,1]$. When $\mathscr{A}$ is the von Neumann algebra
$\mathcal{B}(\mathscr{H})$ of bounded operators on a Hilbert space
$\mathscr{H}$, then `effect' has a consistent meaning, since
$[0,1]_{\mathcal{B}(\mathscr{H})} = \Ef(\mathscr{H})$. Moreover, density operators
$\varrho$ on $\mathscr{H}$ are in one--one correspondence with states
of~$\mathcal{B}(\mathscr{H})$, via $\varrho \mapsto
\tr(\varrho(\,\cdot\,))$; in fact, this correspondence extends to a
linear bipositive isometry between trace-class operators
on~$\mathscr{H}$ and \emph{normal} --- but not necessarily positive
--- functionals on~$\mathcal{B}(\mathscr{H})$
(see~\cite[Thm~2.68]{Alfsen2012}).

For states of von Neumann algebras we use \emph{half} of the operator
norm as distance, since it coincides with the `validity' distance
whose formulation is by now familiar. The proof is again delegated to
the appendix.

\begin{proposition}
\label{prop:vlddist}
Let $\varrho_{1}, \varrho_{2} \colon \mathscr{A} \rightarrow \C$ be
two states of a von Neumann algebra $\mathscr{A}$. Their
\emph{validity distance} $\vld(\varrho_{1}, \varrho_{2})$, as defined
on the left below, satisfies:
\[ \begin{array}{rcccccl}
\vld(\varrho_{1}, \varrho_{2})
& \coloneqq\!\! &
\displaystyle\bigvee_{e\in[0,1]_{\mathscr{A}}}\Big|\,\varrho_{1}\models e -
   \varrho_{2}\models e\,\Big|
& = &
\displaystyle\max\limits_{s\in[0,1]_{\mathscr{A}}\textrm{ sharp}}
   \varrho_{1}\models s - \varrho_{2}\models s
& \,=\, &
\frac{1}{2}\opnorm{\varrho_{1} - \varrho_{2}}.
\end{array} \]

\noindent Via the last equation it is easy to see that $\vld$ is a
complete metric. \QED
\end{proposition}

\begin{corollary}
\label{cor:vNAvld}
Let $\mathscr{A}$ be a von Neumann algebra.
\begin{enumerate}
\item \label{cor:vNAvld:ev} For each predicate $e\in
  [0,1]_{\mathscr{A}}$ the `evaluate at $e$' map $\evmap_{e} = (-)(e)
  = (-) \models e \colon \Stat(\mathscr{A}) \rightarrow [0,1]$ is both
  affine and non-expansive.

\item \label{cor:vNAvld:conv} The convex map $\alpha \colon
  \Dst(\Stat(\mathscr{A})) \rightarrow \Stat(\mathscr{A})$ is
  non-expansive.

\item \label{cor:vNAvld:stat} The `states' functor $\Stat = \Hom(-,\C)
  \colon \op{\vNA} \rightarrow \Conv$ restricts to a functor $\Stat
  \colon \op{\vNA} \rightarrow \ConvCMet$. 
\end{enumerate}
\end{corollary}

\begin{myproof}
\begin{enumerate}
\item It is standard that the map $\evmap_{e}$ is affine, so we
  concentrate on its non-expansiveness: for states $\varrho_{1},
  \varrho_{2}$ we have:
\[ \begin{array}{rcccccl}
\big|\,\evmap_{e}(\varrho_{1}) - \evmap_{e}(\varrho_{2})\,\big|
& = &
\big|\, \varrho_{1} \models e - \varrho_{2} \models e \,\big|
& \leq &
{\displaystyle\bigvee}_{\!a\in[0,1]_{\mathscr{A}}} \big|\, \varrho_{1} \models a - 
   \varrho_{2} \models a \,\big|
& = &
\vld(\varrho_{1}, \varrho_{2}).
\end{array} \]

\item Suppose we have two formal convex combinations $\Omega =
  \sum_{i} r_{i} \ket{\omega_{i}}$ and $\Psi = \sum_{j}
  s_{j}\ket{\varrho_{j}}$ in $\Dst(\Stat(\mathscr{A}))$. The map
  $\alpha \colon \Dst(\Stat(\mathscr{A})) \rightarrow
  \Stat(\mathscr{A})$ is non-expansive since:
$$\begin{array}{rcl}
\vld\big(\alpha(\Omega), \alpha(\Psi)\big)
& = &
{\displaystyle\bigvee}_{e}\, \big|\, 
   (\sum_{i}r_{i}\cdot\omega_{i}) \models e - 
   (\sum_{j}s_{j}\cdot\varrho_{j}) \models e \,\big|
\\
& = &
{\displaystyle\bigvee}_{e}\, \big|\, 
   \sum_{i}r_{i}\cdot\omega_{i}(e) - 
   \sum_{j}s_{j}\cdot\varrho_{j}(e) \,\big|
\\
& = &
{\displaystyle\bigvee}_{e}\, \big|\, 
   \sum_{i}r_{i}\cdot\evmap_{e}(\omega_{i}) - 
   \sum_{j}s_{j}\cdot\evmap_{e}(\varrho_{j}) \,\big|
\\
& = &
{\displaystyle\bigvee}_{e}\, \big|\, 
   \Omega\models \evmap_{e} - 
   \Psi\models \evmap_{e} \,\big|
\\
& \leq &
{\displaystyle\bigvee}_{p \in\Met(\Stat(\mathscr{A}), [0,1])}\, \big|\,
   \Omega\models p - \Psi\models p\,\big|
\\
& \smash{\stackrel{\eqref{eqn:kantorovich}}{=}} &
\kvd\big(\Omega, \Psi\big).
\end{array}$$

\item We have to prove that for a positive unital map
  $f\colon \mathscr{A} \rightarrow \mathscr{B}$ between von Neumann
  algebras the associated state transformer $f_{*} = (-) \after f
  \colon \Hom(\mathscr{B}, \C) \rightarrow \Hom(\mathscr{A},\C)$ is
  affine and non-expansive. The former is standard, so we concentrate
  on non-expansiveness. Let $\varrho_{1},\varrho_{2} \colon
  \mathscr{B} \rightarrow \C$ be states of $\mathscr{B}$. Then:
\[ \begin{array}[b]{rcl}
\vld\big(f_{*}(\varrho_{1}), f_{*}(\varrho_{2})\big)
& = &
{\displaystyle\bigvee}_{e\in[0,1]_{\mathscr{A}}}\, \big|\, 
   f_{*}(\varrho_{1})(e) - f_{*}(\varrho_{2})(e)\,\big|
\\
& = &
{\displaystyle\bigvee}_{e\in[0,1]_{\mathscr{A}}}\, \big|\, 
   \varrho_{1}(f(e)) - \varrho_{2}(f(e))\,\big|
\\
& \leq &
{\displaystyle\bigvee}_{d\in[0,1]_{\mathscr{B}}}\, \big|\, 
   \varrho_{1}(d) - \varrho_{2}(d)\,\big|
\\
& = &
\vld\big(\varrho_{1}, \varrho_{2}\big).
\end{array} \eqno{\QEDbox} \]
\end{enumerate}
\end{myproof}

\section{Distances between effects (predicates)}\label{sec:effects}

There are several closely connected views on what predicates are in a
probabilistic setting. Informally, one can consider fuzzy predicates
$X \rightarrow [0,1]$ on a space $X$, or only the sharp ones $X
\rightarrow \{0,1\}$. Instead of restricting oneself to truth values
in $[0,1]$, one can use $\R$-valued predicates $X \rightarrow \R$,
which are often called `observables'. Alternatively, one can restrict
to the non-negative ones $X \rightarrow [0,\infty)$. There are ways to
  translate between these views, by restriction, or by completion.
  The relevant underlying mathematical structures are: effect modules,
  order unit spaces, and ordered cones. Via suitable restrictions,
  see~\cite[Lem.~13, Thm.~14]{JacobsMF16} for details, the categories of these
  structures are equivalent. Here we choose to use effect modules
  because they capture $[0,1]$-valued predicates, which we consider to
  be most natural. Moreover, there is a standard adjunction between
  effect modules and the convex sets that we have been using in the
  previous section. This adjunction will be explored in the next
  section.

In this section we recall some basic facts from the theory of effect
modules (see~\cite{Jacobs15d,ChoJWW15b,JacobsW15a}), and add a few new
ones, especially related to $\omega$-joins and metric completeness,
see Proposition~\ref{prop:weffmod}. With these results in place, we
observe that in our main examples --- fuzzy predicates on a set and
effects in a von Neumann algebras --- the induced `Archimedean' metric
can also be expressed using validity $\models$, but now in dual form
wrt.\ the previous section: for the distance between two predicates we
now take a join over all states and use the validities of the two
predicates in these states.

We briefly recall what an effect module is, and refer
to~\cite{Jacobs15d} and its references for more details. This involves
three steps.
\begin{enumerate}
\item A \emph{partial commutative monoid} (PCM) is given by a set $E$
  with an element $0\in E$ and a partial binary operation $\ovee
  \colon E \times E \rightarrow E$ which is commutative and
		associative, in a suitably partial sense,
        and has $0$ has unit
  element:  given~$a,b,c\in E$,
the expression
 $a\ovee(b\ovee c)$ is defined
iff $(a\ovee b)\ovee c$ is defined,
and they are equal in that case;
$a\ovee b$ is defined iff $b\ovee a$ is defined,
and they are equal in that case;
and $a\ovee 0$ is always defined, and is equal to~$a$.

\item An \emph{effect algebra} is a PCM $E$ in which each element
  $x\in E$ has a unique orthosupplement $x^{\bot} \in E$ with $x\ovee
  x^{\bot} = 1$, where $1 = 0^{\bot}$. Moreover, if $x \ovee 1$ is
  defined, then $x=0$. Each effect algebra carries a partial order
  given by: $x \leq y$ iff $x \ovee z = y$ for some~$z$. It satisfies
  $x \leq y$ iff $y^{\bot} \leq x^{\bot}$. 
        Moreover, $x\ovee y$ exists
        iff $x\leq y^\perp$ iff $y \leq x^\perp$.
        For more information on
  effect algebras we refer to~\cite{DvurecenskijP00}.

\item An \emph{effect module} is an effect algebra $E$ with a (total)
  scalar multiplication operation $[0,1]\times E \rightarrow E$ which
  acts as a bihomomorphism: it preserves in each coordinate separately
  scalar multiplications $\cdot$ and partial sums $(\ovee,0)$, when
  defined, and maps the pair $(1,1)$ to $1$.
\end{enumerate}

\noindent We write $\EMod$ for the category of effect modules. A map
$f\colon E\rightarrow D$ in $\EMod$ preserves~$1$, sums $\ovee$, when
they exist, and scalar multiplication; such an $f$ then also preserves
orthosupplements and~$0$. There are (non-full) subcategories $\DcEMod
\hookrightarrow \wEMod \hookrightarrow \EMod$ of \emph{directed
  complete} and \emph{$\omega$-complete} effect modules, with joins of
directed (or countable ascending) subsets, with respect to the
existing order of effect algebras. The sum $\ovee$ and scalar
multiplication $\cdot$ operations are required to preserve these joins
in each argument separately\footnote{In fact, it can be shown that
  maps $(-)\ovee y$ preserve directed (or countable ascending) 
  joins automatically when all such directed (or countable ascending)
  joins exist, see
  Lemma~\ref{lem:ard}~\eqref{lem:ard:minus:oveepres}.  Preservation by
  scalar multiplication can also be proved, but is outside the scope
  of this paper.}. Since taking the orthosupplement $a\mapsto a^\perp$
is an order anti-isomorphism it sends joins to meets and vice-versa.
In particular, $\omega$/directed meets exist in $\omega$-/directed
complete effect modules.  Morphisms in $\DcEMod$ and $\wEMod$ are
homomorphisms of effect modules that additionally preserve the
relevant joins.

Below it is shown how this effect module structure arises naturally in
our main examples. The predicate functors $\Pred$ are special cases of
constructions for `effectuses', see~\cite{Jacobs15d}.

\begin{lemma}
\label{lem:predfun}
\begin{enumerate}
\item \label{lem:predfun:Dst} For the distribution monad $\Dst$ on
  $\Sets$ there is a `predicate' functor on its Kleisli category:
\[ \xymatrix{
\Kl(\Dst)\ar[r]^-{\Pred} & \op{\DcEMod} \qquad\mbox{given by}\qquad
   {\left\{\begin{array}{rcl} 
    X & \mapsto & [0,1]^{X} \\ 
   \smash{\big(X \xrightarrow{f}\Dst(Y)\big)} & \;\mapsto\; & 
   \smash{\big([0,1]^{Y} \xrightarrow{f^*} [0,1]^{X}\big)}
   \end{array}\right.}
} \]

\noindent This functor is faithful, and it is full (\& faithful) if we
restrict it to the subcategory $\Klf(\Dst) \hookrightarrow \Kl(\Dst)$
with \emph{finite} sets as objects.

\item \label{lem:predfun:vNA} There is also a `predicate' functor:
\[ \xymatrix{
\op{\vNA}\ar[r]^-{\Pred} & \op{\DcEMod} \qquad\mbox{given by}\qquad
   {\left\{\begin{array}{rcl} 
    \mathscr{A} & \mapsto & [0,1]_{\mathscr{A}} \\ 
   \smash{\mathscr{B} \xrightarrow{f}\mathscr{A}} & \mapsto & 
   [0,1]_{\mathscr{B}} \xrightarrow{f} [0,1]_{\mathscr{A}}
   \end{array}\right.}
} \]

\noindent This functor is full and faithful.
\end{enumerate}
\end{lemma}

Writing $\op{(-)}$ on both sides in point~\eqref{lem:predfun:vNA}
looks rather formal, but makes sense since the category $\vNA$ of von
Neumann algebras is naturally used in opposite form, see also the next
section.

\begin{myproof}
\begin{enumerate}
\item It is easy to see that the set $[0,1]^{X}$ of fuzzy predicate on
  a set $X$ is an effect module, in which a sum $p\ovee q$ exists if
  $p(x) + q(x) \leq 1$ for all $x\in X$, and in that case $(p\ovee
  q)(x) = p(x) + q(x)$.  Clearly, $p^{\bot}(x) = 1 - p(x)$ and
  $(r\cdot p)(x) = r\cdot p(x)$ for a scalar $r\in [0,1]$. The induced
  order on $[0,1]^{X}$ is the pointwise order, which is (directed)
  complete.

For a Kleisli map $f\colon X \rightarrow \Dst(Y)$ the predicate
transformation map $f^{*} \colon [0,1]^{Y} \rightarrow [0,1]^{X}$ 
from~\eqref{eqn:setstransformations} preserves the effect module
structure. Moreover, it is Scott-continuous by the following
argument. Let $q_{i}\in [0,1]^{X}$ be a directed collection of
predicates, and let $x\in X$. Write the support of $f(x)\in\Dst(Y)$
as $\{y_{1}, \ldots, y_{n}\}$. Then:
\[ \begin{array}{rcl}
f^{*}\big(\bigvee_{i} q_{i}\big)(x)
& \smash{\stackrel{\eqref{eqn:setstransformations}}{=}} &
f(x)(y_{1})\cdot \big(\bigvee_{i} q_{i}\big)(y_{1}) + \cdots +
   f(x)(y_{n})\cdot \big(\bigvee_{i} q_{i}\big)(y_{n})
\\
& = &
\big(\bigvee_{i} f(x)(y_{1})\cdot q_{i}(y_{1})\big) + \cdots +
   \big(\bigvee_{i} f(x)(y_{n})\cdot q_{i}(y_{n})\big)
    \\
    &&\qquad\qquad \mbox{since $+$ is Scott-continuous}
\\
& = &
\bigvee_{i} f(x)(y_{1})\cdot q_{i}(y_{1}) + \cdots +
   f(x)(y_{n})\cdot q_{i}(y_{n}) 
\\
& = &
\bigvee_{i} f^{*}(q_{i})(x)
\\
& = &
\big(\bigvee_{i}f^{*}(q_{i})\big)(x).
\end{array} \]

\noindent Assume $f^{*}=g^{*}$ for $f,g\colon X \rightarrow \Dst(Y)$,
and let $x\in X$, $y\in Y$. Write $\indic{\{y\}}\in [0,1]^{Y}$ for the
singleton predicate that is $1$ on $y\in Y$ and zero everywhere
else. Then $f(x)(y) = f^{*}(\indic{\{y\}})(x) =
g^{*}(\indic{\{y\}})(x) = g(x)(y)$. Hence $f=g$, showing that $\Pred$
is faithful.

Now let $X,Y$ be finite sets and $h\colon [0,1]^{Y} \rightarrow
[0,1]^{X}$ be a map in $\DcEMod$. Define $f(x)(y) = h(\indic{\{y\}})(x)
\in [0,1]$. We claim that $f(x)$ is a distribution on $Y = \{y_{1},
\ldots, y_{n}\}$, say, and that $f^{*} = h$. This works as follows.
\[ \begin{array}{rcl}
\qquad\begin{array}{rcl}
\sum_{y\in Y} f(x)(y)
& = &
\sum_{i} h(\indic{\{y_i\}})(x)
\\
& = &
\big(\bigovee_{i} h(\indic{\{y_i\}})\big)(x)
\\
& = &
h\big(\bigovee_{i} \indic{\{y_i\}}\big)(x)
\\
& = &
h(\indic{Y})(x)
\\
& = &
\indic{X}(x)
\\
& = &
1.
\end{array}
& \qquad\qquad &
\begin{array}{rcl}
f^{*}(q)(x)
& = &
\sum_{i} f(x)(y_{i})\cdot q(y_{i})
\\
& = &
\bigovee_{i} h(\indic{\{y_{i}\}})(x) \cdot q(y_{i})
\\
& = &
h(\bigovee_{i} q(y_{i}) \cdot \indic{\{y_{i}\}})(x)
\\
& = &
h(q)(x).
\end{array}
\end{array} \]

\item It is not hard to see that the unit interval
  $[0,1]_{\mathscr{A}}$ of a unital $C^*$-algebra $\mathscr{A}$ is an effect
  module, see also~\cite{Jacobs15d}. If $\mathscr{A}$ is a von Neumann
  algebra, then this interval is a dcpo, by definition. Each map $f$
  of von Neumann algebras restricts to these intervals, and is in fact
  entirely determined by its behaviour on unit intervals: an arbitrary
  element can be written as a linear combination of (four) positive
  elements (see Corollary~4.2.4 of~\cite{Kadison2015}); 
  the latter can be scaled down with a scalar, if needed, so
  that they fit in the unit interval. \QED
\end{enumerate}
\end{myproof}

For comparison with what follows
we recall that the Archimedean property of an order unit space
(see~\cite{nagel1974,kadison1951}) with unit~$1$
is typically formulated as follows. Let $x$ be an arbitrary
element that satisfies $x \leq \frac{1}{n}\cdot 1$, for all $n\geq 1$,
then $x \leq 0$. This Archimedean property is crucial for defining a
norm on order unit spaces.

An analogous Archimedean property is given for effect modules
in~\cite{JacobsM12b,JacobsMF16}. Its formulation is more subtle, and
runs as follows. For arbitrary elements $x,y$, if $\frac{1}{2}\cdot x
\leq \frac{1}{2}\cdot y \ovee \frac{1}{2n}\cdot 1$ for all $n\geq 1$,
then $x\leq y$. This formulation uses the fact that sums $r\cdot x
\ovee s\cdot y$ with $r+s\leq 1$ always exist in an effect module.

Also for Archimedean effect modules one can define an `Archimedean'
distance function $\ard$ as:
\[ \begin{array}{rcl}
\ard(x,y)
& = &
\max\Big(\bigwedge\setin{r}{(0,1]}{\frac{1}{2}\cdot x \leq 
   \frac{1}{2}\cdot y \ovee \frac{r}{2}\cdot 1}, \;
   \bigwedge\setin{r}{(0,1]}{\frac{1}{2}\cdot y \leq 
   \frac{1}{2}\cdot x \ovee \frac{r}{2}\cdot 1}\Big)
\end{array} \]

\noindent In this situation we can write $\|x\| = \ard(0,x) \in
          [0,1]$, so that $x \leq \|x\|\cdot 1$
          (see Lemma~\ref{lem:ard}\eqref{lem:ard:ard} below). But we need to be
          careful that we cannot express the distance $\ard$ in terms
          of $\|-\|$ since there is no general subtraction in effect
          modules --- but there is a partial operation $\ominus$, see
          below.

In~\cite{JacobsM12b,JacobsMF16} it is shown that:
\begin{itemize}
\item the full subcategory $\AEMod$ of Archimedean effect modules is
  equivalent to the category of order unit spaces; the `Archimedean'
  distances on order unit spaces and effect modules coincide;

\item Archimedean effect modules carry this (1-bounded) metric $\ard$,
  and all maps of effect modules are automatically non-expansive. This
  gives a functor $\AEMod \rightarrow \Met$.
\end{itemize}

We need to collect a few basic facts about this Archimedean distance
function $\ard$, especially about its relation to (partial)
subtraction $\ominus$ in the last point below.

\begin{lemma}
\label{lem:ard}
Let $E$ be an Archimedean effect module. For $x,y\in E$ with $x\leq y$
    one can define\footnote{Indeed,
    recall that~$a\ovee b$ exists iff~$b\leq a^\perp$.
    Thus $y^\perp\ovee x$ exists since  $x\leq y\equiv y^{\perp\perp}$.}
     $y\ominus x = (y^{\bot} \ovee x)^{\bot}$. Then:
\begin{enumerate}
\item \label{lem:ard:minus} This minus operation $\ominus$ satisfies
  the following properties:
\begin{enumerate}
\item $x \ominus 0 = x$ and $1 \ominus y = y^{\bot}$ and $x \ominus x = 0$;

\item \label{lem:ard:minus:eqn} if $y\leq z$ then: $x \ovee y = z$ iff
  $x = z \ominus y$; in particular, $x = (x \ovee y) \ominus y$ and
  $(z \ominus y) \ovee y = z$;

\item \label{lem:ard:minus:oveeleq} $x \ovee y \leq z$ iff $x \leq z
    \ominus y$ (and $y\leq z$);

\item \label{lem:ard:minus:minusfromsum} if $x\leq y$ then $(y\ovee z)
  \ominus x = (y\ominus x) \ovee z$;

\item \label{lem:ard:minusleq} if $x\leq y\leq z$ then $y\ominus x
  \leq z \ominus x$;

\item \label{lem:ard:minusleqovee} if $x \geq y$ then $x \leq y\ovee
  z$ iff $x\ominus y \leq z$;

\item \label{lem:ard:minus:half} if $x \leq y$ then $r\cdot
  y \ominus r\cdot x = r\cdot (y \ominus x)$ for $r\in [0,1]$;

\item \label{lem:ard:minus:scalar} if $r \leq s$ in $[0,1]$, 
	which is itself an Archimedean effect module, then $s
  \ominus r = s - r$ and $(s-r)\cdot x = s\cdot x \ominus r\cdot x$;

\item \label{lem:ard:minus:oveepres} 
    Let~$S$ be a non-empty subset of~$[0,y^\perp]_E$.
If~$S$ has a join~$\bigvee S$ in~$E$,
then~$y\ovee \bigvee S$
exists, and
        is the join of $\set{y\ovee s}{s\in S}$
        in~$E$.
If~$S$ has a meet~$\bigwedge S$ in~$E$,
        and\footnote{The condition that
        the $y\ovee s$ have a meet in~$E$ cannot be dropped.
        To see this, 
        recall from~\cite{kadison1951order}, Lemma~2, that
        projections~$P$ and~$Q$ on closed linear subspaces
        $C$ and~$D$ of a Hilbert sapce~$\mathscr{H}$, respectively,
        have an infimum
        in the set of positive operators $\mathscr{B}(\mathscr{H})_+$
        on~$\mathscr{H}$,
        namely the projection~$R$ onto~$C\cap D$.
        However, by~\cite{kadison1951order}, Corollary~4, 
        $P$ and~$Q$ only have an infimum
        in the space of self-adjoint bounded operators
        $\mathscr{B}(\mathscr{H})_{\mathrm{sa}}$ on~$\mathscr{H}$
        when~$P$ and~$Q$ commute.
        By inspecting and adapting the proofs of these results,
        one easily sees that when~$P$ and~$Q$
        do not commute,
        then $\frac{1}{2}P$ and~$\frac{1}{2}Q$
        have~$\frac{1}{2}R$ as meet in~$[0,1]_{\mathscr{B}(\mathscr{H})}$,
        while $\frac{1}{2}\cdot 1\ovee \frac{1}{2}P$
        and~$\frac{1}{2}\cdot 1 \ovee \frac{1}{2}Q$
        have no meet in~$[0,1]_{\mathscr{B}(\mathscr{H})}$ at all.}
        $\set{y\ovee s}{s\in S}$
        has a meet~$\bigwedge_{s\in S} y\ovee s$
        in~$E$,
        then~$y\ovee \bigwedge S=
        \bigwedge_{s\in S} y\ovee s$.
\item \label{lem:ard:minus:ominuspres} 
    Let~$S$ be a non-empty
        subset of~$[y,1]_E$.
If~$S$ has a join~$\bigvee S$ in~$E$,
        and $\set{s\ominus  y}{s\in S}$
        has a join $\bigvee_{s\in S} s\ominus y$
        in~$E$,
        then $\bigvee_{s\in S} s\ominus y = (\bigvee S)\ominus y$.
If~$S$ has a meet~$\bigwedge S$ in~$E$,
        then~$(\bigwedge S) \ominus y$ exists,
        and is the meet of~$\set{s\ominus y}{s\in S}$
        in~$E$.
\end{enumerate}
\item \label{lem:ard:minus:scalarpres}
Scalar multiplication preserves meets in its first argument:
		$(\bigwedge S)\cdot 1
		= \bigwedge_{s\in S} (s\cdot 1)$
		for any set of scalars $S\subseteq[0,1]$.

\item \label{lem:ard:ard} Given $r\in [0,1]$
and~$x\leq y$ from~$E$
we have $\ard(x,y) \leq r$
iff  $y\ominus x \leq
r \cdot 1$.
In particular, $y\ominus x \leq \ard(x,y)\cdot 1$.
Moreover,
$\|y\| \leq r$
iff $y \cdot 1 \leq r$;
and~$y\cdot 1 \leq \|y\|$.

\item \label{lem:ard:cont} The sum $\ovee$
is continuous  wrt.\ the Archimedean
metric~$\ard$, in the sense that when
$x_1,x_2,\dotsc \in E$
converge to~$x\in E$ wrt.\ $\ard$,
and $y_1,y_2,\dotsc \in E$
converge to~$y\in E$,
and $x_n$ is summable with~$y_n$ for each~$n$,
then $x\ovee y$ exists too,
and~$\ard(x_n\ovee y,x\ovee y)\to 0$.
Orthosupplement
  $(-)^{\bot}$, scalar multiplication $\cdot$,
and~$\ominus$ are continuous in a similar sense too.

\end{enumerate}
\end{lemma}

\begin{myproof}
\begin{enumerate}
\item The first point is trivial, and left to the
  reader. For~\eqref{lem:ard:minus:eqn} we use: $x = z \ominus y =
  (z^{\bot} \ovee y)^{\bot}$ iff $x^{\bot} = z^{\bot} \ovee y$ iff $x
  \ovee z^{\bot} \ovee y = 1$ iff $z = x \ovee y$. Next,
  for~\eqref{lem:ard:minus:oveeleq},
\[ \begin{array}{rcl}
x \ovee y \leq z
\hspace*{\arraycolsep}\Leftrightarrow\hspace*{\arraycolsep}
\ex{w}{x \ovee y \ovee w = z}
& \Leftrightarrow &
\ex{w}{x \ovee w = z \ominus y} \qquad \mbox{as just shown}
\\
& \Leftrightarrow &
x \leq z \ominus y.
\end{array} \]

\noindent Point~\eqref{lem:ard:minus:minusfromsum} is obtained as
follows. We have:
\[ \begin{array}{rcccl}
(y\ovee z)^{\bot} \ovee x \ovee (y\ominus x) \ovee z
& \smash{\stackrel{\eqref{lem:ard:minus:eqn}}{=}} &
(y\ovee z)^{\bot} \ovee y\ovee z
& = &
1,
\end{array} \]

\noindent so that $(y\ominus x) \ovee z = \big((y\ovee z)^{\bot} \ovee
x\big)^{\bot} = ((y\ovee z) \ominus x$.

For~\eqref{lem:ard:minusleq} let $x\leq y\leq z$, say via $z = y \ovee
w$. Then $z \ominus x = (y\ovee w)\ominus x = (y\ominus x) \ovee w$ by
the previous point. Hence $y\ominus x \leq z \ominus x$.

Assume now $x \geq y$ for~\eqref{lem:ard:minusleqovee}. In one
direction, if $x \leq y\ovee z$, then, by the previous point, $x
\ominus y \leq (y\ovee z) \ominus y = z$. The other direction follows
similary by adding $y$ on both sides.

For~\eqref{lem:ard:minus:half} let $x\leq y$ and $r\in [0,1]$.
        Then~$r\cdot y = r\cdot (x\ovee (y\ominus x))
        = (r\cdot x)\ovee(r\cdot (y\ominus x))$,
        so $r\cdot x\leq r\cdot y$, and
        $(r\cdot y)\ominus (r\cdot x) = r\cdot (y\ominus x)$,
        by~\eqref{lem:ard:minus:eqn}.

\auxproof{
\noindent Clearly, for $r \leq s$ in $[0,1]$ we have:
\[ \begin{array}{rcccl}
s \ominus r
& = &
1 - \big((1-s) + r\big)
& = &
s - r.
\end{array} \]

\noindent We obtain $(s-r)\cdot x = s\cdot x \ominus r\cdot x$ from: 
\[ \begin{array}{rcccl}
(s-r)\cdot x \ovee r\cdot x
& = &
s\cdot x
& = &
\big(s\cdot x \ominus r\cdot x\big) \ovee r\cdot x.
\end{array} \]
}

Point~\eqref{lem:ard:minus:scalar} is easy and left
to the reader.  
For~\eqref{lem:ard:minus:oveepres}
and~\eqref{lem:ard:minus:ominuspres}
first note 
that the map  $y\ovee (-)\colon [0,y^\perp]_E
\longrightarrow [y,1]_E$
is not only order preserving,
but also an order isomorphism,
with inverse  $(-)\ominus y\colon [y,1]_E\longrightarrow [0,y^\perp]_E$,
by~\eqref{lem:ard:minus:eqn} and~\eqref{lem:ard:minusleq}.
Therefore
$y\ovee (-)\colon [0,y^\perp]_E
\longrightarrow [y,1]_E$
preserves and reflects joins.

So if a non-empty subset~$S$ of $\subseteq [0,y^\perp]_E$
has a join~$\bigvee S$ in~$E$
(which must be the join in~$[0,y^\perp]_E$ too,)
then~$y\ovee \bigvee S$
is the join of the~$y\ovee s$
in~$[y,1]_E$.
We claim that~$y\ovee \bigvee S$
is the join of the~$y\ovee s$
in~$E$ too,
using here that~$S$ is non-empty.
Indeed, let~$u\in E$ with $y\ovee s\leq u$
for all~$s\in S$ be given;
we must show that~$y\ovee \bigvee S\leq u$.
Since there is some~$s_0\in S$,
we have $y\leq y\ovee s_0\leq u$,
and so~$u\in [y,1]_E$,
which entails that~$y\ovee \bigvee S\leq u$,
since $y\ovee \bigvee S$ is the least upper bound
of the~$y\ovee s$ in~$[y,1]_E$.

Now suppose that~$S$ is a non-empty
subset of~$[0,y^\perp]_E$
that has a meet in~$E$,
and suppose that $\set{y\ovee s}{s\in S}$
has a meet $\bigwedge_{s\in S} y\ovee s$ \emph{in~$E$} too.
We must show that 
$\bigwedge_{s\in S} y\ovee s = y\ovee \bigwedge S$.
(Note that $\bigwedge S\leq y^\perp$
since~$S$ is non-empty,
and so~$y\ovee \bigwedge S$
exists.)
Since
$y\ovee (-)\colon [0,y^\perp]_E
\longrightarrow [y,1]_E$,
being an order isomorphism,
preserves (and reflects) meets,
and~$\bigwedge S$ is the meet of~$S$
in~$E$, and so
in~$[0,y^\perp]_E$ too,
we see that~$y\ovee \bigwedge S$
is the meet of the~$y\ovee s$ in~$[y,1]_E$.
Since~$\bigwedge_{s\in S} y\ovee s$
is the meet of the~$y\ovee s$ in~$E$, and thus in~$[y,1]_E$ too,
we get $y\ovee \bigwedge S =\bigwedge_{s\in S} y\ovee s$.

Whence~\eqref{lem:ard:minus:oveepres} holds,
and~\eqref{lem:ard:minus:ominuspres}
is established similarly.

\auxproof{
If $x_{i} \orthogonal y$, then $\bigwedge_{i}x_{i} \leq x_{j} \leq y^{\bot}$,
so that $\bigwedge x_{i} \orthogonal y$. Moreover,
\[ \begin{array}{rcll}
z \leq (\bigwedge x_{i}) \ovee y
& \Longleftrightarrow &
z \ominus y \leq \bigwedge x_{i} & \mbox{by~\eqref{lem:ard:minusleqovee}}
\\
& \Longleftrightarrow &
z \ominus y \leq x_{i}, \mbox{ for all $i$}
\\
& \Longleftrightarrow &
z \leq x_{i} \ovee y, \mbox{ for all $i$}
\\
& \Longleftrightarrow &
z \leq \bigwedge x_{i} \ovee y.
\end{array} \]
}

\hide{%

Finally, for~\eqref{lem:ard:minus:scalarpres}, let $(-)\cdot x$
preserve joins. Then:
\[ \begin{array}{rcll}
\big(\bigwedge r_{i}\big) \cdot x
\hspace*{\arraycolsep}=\hspace*{\arraycolsep}
\big(\bigvee r_{i}^{\bot}\big)^{\bot} \cdot x
& = &
\big(1 - \bigvee r_{i}^{\bot}\big) \cdot x
\\
& = &
1\cdot x \ominus \big(\bigvee r_{i}^{\bot}\big) \cdot x
   \qquad & \mbox{by~\eqref{lem:ard:minus:scalar}}
\\
& = &
1\cdot x \ominus \big(\bigvee r_{i}^{\bot} \cdot x\big)
\\
& = &
\bigwedge 1\cdot x \ominus r_{i}^{\bot} \cdot x
   & \mbox{by~\eqref{lem:ard:minus:ominuspres}}
\\
& = &
\bigwedge r_{i}\cdot x.
\end{array} \]

\noindent Next let $r\cdot (-)$ preserve joins. We use
that $r\cdot (x^{\bot}) = (r\cdot x \ovee r^{\bot}\cdot 1)^{\bot}$ in:
\[ \begin{array}{rcl}
r\cdot \big(\bigwedge x_{i}\big)
\hspace*{\arraycolsep}=\hspace*{\arraycolsep}
r \cdot \big(\bigvee x_{i}^{\bot}\big)^{\bot}
& = &
\Big(r \cdot \big(\bigvee x_{i}^{\bot}\big) \ovee r^{\bot}\cdot 1\Big)^{\bot}
\\
& = &
\Big(\bigvee r \cdot x_{i}^{\bot} \ovee r^{\bot}\cdot 1\Big)^{\bot}
\\
& = &
\bigwedge r\cdot x_{i}.
\end{array} \] 

} 


\item
Given a set~$S\subseteq [0,1]$ of scalars,
we must show that $(\bigwedge S)\cdot 1
= \bigwedge_{s\in S} s\cdot 1$.
The difficulty here is not whether
$(\bigwedge S)\cdot 1$ is a lower bound
of the $s\cdot 1$,
but whether it is the greatest lower bound.
To prove this,
let~$x\in E$ with $x\leq s\cdot 1$ for all~$s\in S$ be given;
we must prove that~$x\leq (\bigwedge S)\cdot 1$.
Since~$E$ is Archimedean it suffices
(by the definition of the Archimedean property for
effect module above)
to prove that for given $n>1$ we have 
$\frac{1}{2} \cdot x \leq 
\frac{1}{2}\cdot (\,(\bigwedge S)\cdot 1\,)
\ovee\frac{1}{2n}\cdot 1\equiv 
\frac{1}{2} (\bigwedge S + \frac{1}{n}) \cdot 1$.
Since the elements of~$S$ are just plain real numbers
we can find~$s\in S$ with $s\leq \bigwedge S+\frac{1}{n}$.
Using this~$s$, we see that~$\frac{1}{2}\cdot x \leq (\frac{1}{2} s) \cdot 1
\leq \frac{1}{2}(\bigwedge S + \frac{1}{n})\cdot 1$.
Whence~$ x \leq (\bigwedge S)\cdot 1$.

\item 
Let sequences~$x_1,x_2,\dotsc$ 
and~$y_1,y_2,\dotsc$
in~$E$
$\ard$-converging to elements~$x$ and~$y$ of~$E$, respectively,
be given,
such that~$x_n\ovee y_n$ exists
for all~$n$.
We must show that~$x\ovee y$ exists,
and that $x_1\ovee y_1,\ x_2\ovee y_2,\  \dotsc$ $\ard$-converges to~$x\ovee y$.

To show that~$x\ovee y$ exists
we need to show that~$x\leq y^\perp$,
and for this in turn,
it suffices
 given integer $n>0$
(since~$E$ is Archimedean)
to prove that~$\frac{1}{2}x\leq 
\frac{1}{2}y^\perp \ovee \frac{1}{2n}\cdot 1$.
Since~$\ard(x_m,x)\to 0$ as $m\to\infty$,
we can find an~$M$
such that~$\ard(x_m,x)<\frac{1}{2n}$
for all~$m\geq M$.
From this,
and
the definition of~$\ard$,
it follows readily that $\frac{1}{2} x \leq \frac{1}{2} x_m 
\ovee \frac{1}{4n}\cdot 1$
for all~$m\geq M$.
By a similar argument, but now
using that~$\ard(y_m,y)\to 0$ as~$m\to\infty$,
we can, by choosing~$M$ larger if necessary,
have $\frac{1}{2} y \leq \frac{1}{2} y_m \ovee
\frac{1}{4n}\cdot 1$ for all~$m\geq M$ too.
Note that
$(\frac{1}{2}a\ovee \frac{1}{2}b)^\perp = \frac{1}{2}a^\perp \ovee
\frac{1}{2}b^\perp$
for all~$a,b\in E$.
So upon application of~$(\,\cdot\,)^\perp$,
the aforementioned inequality gives
\begin{alignat*}{3}
\textstyle
\frac{1}{2}y_m^\perp \,\ovee\, \frac{1}{2}(\frac{1}{2n}\cdot 1)^\perp
    \ &\textstyle= \  (\frac{1}{2} y_m\ovee \frac{1}{4n}\cdot 1)^\perp 
    \\
    \ &\textstyle\leq\ (\frac{1}{2} y \ovee \frac{1}{2} \cdot 1^\perp)^\perp
    \\
    \ &\textstyle= \ \frac{1}{2} y^\perp \,\ovee\, \frac{1}{2}\cdot 1
\ = \  \frac{1}{2} y^\perp 
    \,\ovee\, \frac{1}{2}(\frac{1}{2n}\cdot 1) 
    \,\ovee\, \frac{1}{2}(\frac{1}{2n}\cdot 1)^\perp,
\end{alignat*}
which implies that
$\frac{1}{2}y_m^\perp \leq \frac{1}{2}y^\perp\ovee \frac{1}{4n}\cdot 1$,
for all~$m\geq M$. 
As the final ingredient,
note that~$x_M\leq y_M^\perp$
since $x_M\ovee y_M$ exists.
Altogether we get:
$$\textstyle
    \frac{1}{2}x\ \leq\ \frac{1}{2} x_M\ovee \frac{1}{4n}\cdot 1
\\
    \leq\  \frac{1}{2}y_M^\perp \ovee \frac{1}{4n} \cdot 1
    \\ 
    \leq\  \frac{1}{2}y^\perp \ovee \frac{1}{2n}\cdot 1.
    $$
Whence~$x\leq y^\perp$, and so~$x\ovee y$ exists.

Concerning the continuity of~$\ovee$ 
it remains to be shown that~$x_n\ovee y_n$
converges to~$x\ovee y$.
For this we need the
observation that $\ard(a\ovee c,b\ovee c)=\ard(a,b)$
for all $a,b,c\in E$ for which~$a\ovee c$ and~$b\ovee c$ exist.
(Hint: looking at the definition of~$\ard$
note that given $r\in (0,1]$ we
have $\frac{1}{2} a \leq \frac{1}{2}b\ovee \frac{1}{2r}\cdot 1$
iff $\frac{1}{2} (a \ovee c)
\leq \frac{1}{2} (b \ovee c) \ovee \frac{1}{2r}\cdot 1$.)
Indeed, this identity gives us
$\ard(x_n\ovee y_n,x\ovee y)
\leq \ard(x_n\ovee y_n, x_n\ovee y)
+ \ard(x_n\ovee y, x\ovee y)
= \ard(y_n,y) + \ard(x_n,x)$,
and so~$\ard(x_n\ovee y_n,x\ovee y)\to 0$
as~$n\to \infty$.

The continuity of~$\cdot$ and~$(\,\cdot\,)^\perp$
follows along similar lines,
but involves the equations
$\ard(x^\perp,y^\perp)=\ard(x,y)$,
$\ard(r\cdot x, r\cdot y)=r\cdot\ard(x,y)$
and $\ard(r\cdot x, s\cdot x)=\left|r-s\right|\cdot\left\|x\right\|$,
whose proofs we leave to the reader.

\item Let $x\leq y$ in~$E$ and $r'\in [0,1]$ be given.
    Recall that we must show that $y\ominus x\leq r'\cdot 1$
    iff $\ard(x,y)\leq r'$.
Since $x\leq y$,
we have $\bigwedge\set{r\in(0,1]}{\frac{1}{2}\cdot x \leq
  \frac{1}{2}\cdot y \ovee \frac{r}{2}\cdot 1} = 0$, so:
\begin{equation}
    \label{eq:ard:ard}
\textstyle
\ard(x,y)\ = \ \bigwedge\set{r\in (0,1]}{\frac{1}{2}\cdot y
\leq \frac{1}{2}\cdot x \ovee \frac{r}{2}\cdot 1}
\end{equation}
Suppose that~$\ard(x,y)\leq r'$.
We must show that~$y\ominus x \leq r'\cdot 1$.
It suffices to show that~$y\ominus x \leq \ard(x,y)\cdot 1$. Indeed:
\[ \begin{array}[b]{rcl}
\ard(x,y)\cdot 1
& = &
    \big(\bigwedge\set{r\in(0,1]}{\frac{1}{2}\cdot y \leq
    \frac{1}{2}\cdot x \ovee \frac{r}{2}\cdot 1}\big)\cdot 1
\\
& \smash{\stackrel{\eqref{lem:ard:minus:scalarpres}}{=}} &
\bigwedge\set{r\cdot 1}{\frac{1}{2}\cdot y \leq
   \frac{1}{2}\cdot x \ovee \frac{r}{2}\cdot 1}
\\
& \smash{\stackrel{\eqref{lem:ard:minusleqovee}}{=}} &
\bigwedge\set{r\cdot 1}{\frac{1}{2}\cdot y \ominus \frac{1}{2}\cdot x \leq
   \frac{r}{2}\cdot 1}
\\
& \geq &
\big(\frac{1}{2}\cdot y \ominus \frac{1}{2}\cdot x\big) \ovee
   \big(\frac{1}{2}\cdot y \ominus \frac{1}{2}\cdot x\big)
\\
& \smash{\stackrel{\eqref{lem:ard:minus:half}}{=}} &
\frac{1}{2}\cdot \big(y \ominus x\big) \ovee
   \frac{1}{2}\cdot \big(y \ominus x\big)
\\
& = &
y \ominus x.
\end{array}  \]
For the other direction,
suppose that~$y \ominus x \leq r'\cdot 1$.
We must show that $\ard(x,y)\leq r'$.
If~$r'=0$, then this is clearly true 
(since then~$y\ominus x=0$, 
thus $x=y$, thus $\ard(x,y)=0$.)
So we may assume that~$r'\neq 0$.
Then $\frac{1}{2}\cdot y\ominus\frac{1}{2}\cdot x
= \frac{1}{2}\cdot (y\ominus x)
\leq \frac{r'}{2}\cdot 1$,
so $\frac{1}{2}\cdot y \leq \frac{1}{2}\cdot x \ovee \frac{r'}{2}\cdot 1$,
which implies~$\ard(x,y)\leq r'$,
by~\eqref{eq:ard:ard}.
\qedhere
\end{enumerate}
\end{myproof}

\begin{proposition}
\label{prop:weffmod}
Let $E$ be an $\omega$-complete effect module. Then:
\begin{enumerate}
\item \label{prop:weffmod:arch} $E$ is Archimedean;

\item \label{prop:weffmod:complete} $E$ is metrically complete for the
  above Archimedean distance function $\ard$;

\item \label{prop:weffmod:suplim} for each ascending sequence $e_{1}
  \leq e_{2} \leq e_{3} \leq \cdots$ which is Cauchy, one has $\bigvee
  e_{n} = \lim e_{n}$.
\end{enumerate}
\end{proposition}

\begin{myproof}
\begin{enumerate}
\item Assume $\frac{1}{2}\cdot x \leq \frac{1}{2}\cdot y \ovee
  \frac{1}{2n}\cdot 1$ for all $n\geq 1$. We need to prove $x \leq
  y$. Recall that the partial addition
        and scalar multiplication
        preserve all $\omega$-joins,
        by our definition of $\omega$-completeness.
So since~$\bigwedge_n \frac{1}{2n}=0$,
we compute 
$$
\textstyle
        \frac{1}{2}\cdot y
        \ =\ 
        \frac{1}{2}\cdot y
        \ovee \bigwedge_n \frac{1}{2n}\cdot 1
        \ = \ 
        \bigwedge_n \frac{1}{2}\cdot y\ovee \frac{1}{2n}\cdot 1
        \ \geq \ \frac{1}{2}\cdot x.
$$
        Thus~$x = \frac{1}{2}\cdot x \ovee \frac{1}{2}\cdot x
        \leq \frac{1}{2}\cdot y\ovee\frac{1}{2}\cdot y = y$.

\item We use an auxiliary result
	that we will prove in a moment:
\[ \begin{array}{c}
\mbox{assume that for each sequence $a_{1}, a_{2}, \ldots \in
  E$ for which $\sum_{n} \|a_{n}\| \leq 1$,}
\\
\mbox{the sums $b_{N} \coloneqq \bigovee_{n\leq N} a_{n}$ converge;} 
\\ 
\mbox{then $E$ is complete.}
\end{array}\eqno{(*)} \]

\noindent We first remark that the sums $b_{N}:=\ovee_{n\leq N}a_n$
exists,
as can be seen using induction.
Indeed, if~$b_N\equiv \bigovee_{n\leq N}a_n$
		exists, then so does $(\bigovee_{n\leq N}a_n)\,\ovee\,a_{N+1}$,
because 
		since $\sum_{n} \|a_n\|\leq 1$,
		we have 
        $\sum_{n\leq N} \|a_n\|\leq \|a_{N+1}\|^\perp \equiv 1-\|a_{N+1}\|$, 
and thus,
        using Lemma~\ref{lem:ard}\eqref{lem:ard:ard},
$\bigovee_{n\leq N} a_{n} \leq \bigovee_{n\leq N} \|a_{n}\|\cdot 1 =
\big(\sum_{n\leq N} \|a_{n}\|\big) \cdot 1
		\leq \|a_{N+1}\|^\perp\cdot 1
		\leq a_{N+1}^\perp$.

We start by proving that~$E$ is complete using
statement~$(*)$. Let $x_{1}, x_{2}, \ldots \in
E$ be a Cauchy sequence; we need to prove that it converges, given the
assumption in~$(*)$. We replace~$x_1,x_2,\dotsc$ by $\frac{1}{2}\cdot
x_1,\,\frac{1}{2}\cdot x_2,\,\dotsc$ so that we may assume that
$x_n\leq\frac{1}{2}\cdot 1$ for all~$n$, because if~$(\,\frac{1}{2}\cdot
x_n\,)_n$ converges, then so does~$(x_n)_n$,
and since~$x_1,x_2,\dotsc$
is Cauchy, so is $\frac{1}{2}\cdot x_1,\, \frac{1}{2}\cdot x_2,
\,\cdots$. Similarly, by replacing
$(x_n)_n$ by an appropriate subsequence we may assume that
$\ard(x_{m},x_n) < (\frac{1}{2})^{n}$ for all~$m\geq n$.  In
particular, $\ard(x_{n+1},x_n)< (\frac{1}{2})^{n+1}$, 
which implies, by the definition of~$\ard$, that
\[ \begin{array}{rclcrcl}
	\frac{1}{2}\cdot x_{n}
& \leq & 
	\frac{1}{2}\cdot x_{n+1}\,\ovee\,\frac{1}{2} (\frac{1}{2})^{n+1}\cdot 1
& \quad\mbox{and}\quad &
	\frac{1}{2}\cdot x_{n+1}
& \leq & 
	\frac{1}{2}\cdot x_{n}\,\ovee\, \frac{1}{2}(\frac{1}{2})^{n+1}\cdot 1.
\end{array} \]
Since~$x_n\leq \frac{1}{2}\cdot 1$
and $x_{n+1}\leq \frac{1}{2}\cdot 1$,
this implies:
\[ \begin{array}{rclcrcl}
x_{n}
& \leq & 
x_{n+1}\ovee (\frac{1}{2})^{n+1}\cdot 1
& \qquad\mbox{and}\qquad &
x_{n+1}
& \leq & 
x_{n}\ovee (\frac{1}{2})^{n+1}\cdot 1.
\end{array} \]

\noindent We then have:
\[ \begin{array}{rcl}
x_{1}
\hspace*{\arraycolsep}\leq\hspace*{\arraycolsep}
x_{2} \ovee (\frac{1}{2})^{2}\cdot 1
& \leq &
x_{3} \ovee (\frac{1}{2})^{3}\cdot 1 \ovee (\frac{1}{2})^{2} \cdot 1
\\
& \leq & 
\cdots \\
& \leq &
x_{n+1} \ovee (\frac{1}{2})^{n+1}\cdot 1 \ovee \cdots \ovee 
   (\frac{1}{2})^{2}\cdot 1
\\
& = &
x_{n+1} \ovee (\frac{1}{2} - (\frac{1}{2})^{n+1})\cdot 1.
\end{array} \]


The trick is to consider the elements $a_n \coloneqq (x_{n+1} \ovee
(\frac{1}{2})^{n+1}\cdot 1) \ominus x_n$. We check that these $a_{n}$
satisfy the requirement in~$(*)$:
\[ \begin{array}{rcccccl}
a_{n}
& = &
(x_{n+1}\ovee (\frac{1}{2})^{n+1}\cdot 1) \ominus x_n 
& \leq &
(x_n\ovee (\frac{1}{2})^{n+1}\cdot 1\ovee (\frac{1}{2})^{n+1}\cdot 1)\ominus x_n
& = &
(\frac{1}{2})^{n}\cdot 1.
\end{array} \]

\noindent Thus we have $\|a_n\|\leq (\frac{1}{2})^{n}$ (by
Lemma~\ref{lem:ard}\eqref{lem:ard:ard}), and so~$\sum_n
\|a_n\| \leq 1$. We may now additionally assume that the sums $b_{N}
\coloneqq \bigovee_{n\leq N} a_{n}$ converge. These sums can be
re-organised as:
\[ \begin{array}{rcl}
b_{N} 
& = &
\bigovee_{n\leq N} a_{n}
\\
& = &
\big((x_{N+1} \ovee (\frac{1}{2})^{N+1}\cdot 1) \ominus x_{N}\big) \ovee
   \big((x_{N} \ovee (\frac{1}{2})^{N})\cdot 1 \ominus x_{N-1}\big) \ovee \cdots 
\\
& & \qquad \ovee\;
   \big((x_{2} \ovee (\frac{1}{2})^{2})\cdot 1 \ominus x_{1}\big)
\\
& = &
\big(x_{N+1} \ovee (\frac{1}{2})^{N+1}\cdot 1 \ovee (\frac{1}{2})^{N}\cdot 1
   \ovee \cdots \ovee (\frac{1}{2})^{2}\cdot 1\big)
   \ominus x_{1}
\\
& = &
\big(x_{N+1}\ovee (\frac{1}{2}-(\frac{1}{2})^{N+1})\cdot 1\big)\ominus x_1.
\end{array} \]

\noindent We claim that we can now also show that the sequence of
$x_{N}$ converges, since:
\[ \begin{array}{rcl}
x_{N+1}
& = &
(b_{N} \ovee x_1)\ominus (\frac{1}{2}-(\frac{1}{2})^{N+1})\cdot 1.
\end{array} \]

\noindent Indeed, the right-hand-side converges, as $N$ goes to infinity,
by Lemma~\ref{lem:ard}\eqref{lem:ard:cont}.

We will now prove~$(*)$.
So let~$a_1,a_2,\dotsc \in E$ for which
$s \coloneqq \sum_n \|a_n\|\leq 1$ and sums $b_N \coloneqq
\bigovee_{n\leq N} a_n$ exist. These $b_{N}$ form an ascending chain,
so by $\omega$-completeness of $E$, the suppremum $b \coloneqq
\bigvee_{N} b_{N}$ exists. We are done if we can show that $b$ is the
limit of the $b_{N}$.
For $M \leq N$ we have:
\[ \begin{array}{rcl}
b_{N} \ominus b_{M}
\hspace*{\arraycolsep}=\hspace*{\arraycolsep}
a_{N} \ovee \cdots \ovee a_{M+1} 
& \leq &
\|a_{N}\|\cdot 1 \ovee \cdots \ovee \|a_{M+1}\|\cdot 1 
\\
& = &
\big(\|a_{N}\| + \cdots + \|a_{M+1}\|\big)\cdot 1.
\end{array} \]

\noindent This means:
\[ \begin{array}{rcl}
b \ominus b_{M}
& = &
\big(\bigvee_{N}b_{N}\big)\ominus b_{M}
\\
& = &
\big(\bigvee_{N\geq M}b_{N}\big)\ominus b_{M}
\\
& = &
\bigvee_{N\geq M} b_{N}\ominus b_{M}
   \qquad \mbox{by Lemma~\ref{lem:ard}~\eqref{lem:ard:minus:oveepres}}
\\
& \leq &
\bigvee_{N\geq M} \big(\|a_{N}\| + \cdots + \|a_{M+1}\|\big)\cdot 1
\\
& = &
\big(\bigvee_{N\geq M} \|a_{N}\| + \cdots + \|a_{M+1}\|\big)\cdot 1
\\
& = &
\big(s - (\|a_{M}\| + \cdots + \|a_{1}\|)\big)\cdot 1,
  \qquad \mbox{where, recall, $s \coloneqq \sum_n \|a_n\| \in [0,1]$.}
\end{array} \]

\noindent The latter scalar becomes arbitrarily small as $M$ goes
to infinity.
This means that $\ard(b, b_{M})$ can be made
arbitrarily small (see Lemma~\ref{lem:ard}\eqref{lem:ard:ard}.)
Hence $\lim_{M} b_{M} = b$, as required. 

\item Let $e_{1} \leq e_{2} \leq \cdots$ be a Cauchy sequence and let
  $\epsilon > 0$. We can find an $N\in\NNO$ such that $\ard(e_{n},
  e_{m}) < \epsilon$ for all $n,m \geq N$. For $m\geq N$ we have
  $e_{m} \leq \bigvee_{n} e_{n}$, so that:
\[ \begin{array}{rcll}
(\bigvee_{n} e_{n}) \ominus e_{m}
& = &
(\bigvee_{n\geq m} e_{n}) \ominus e_{m}
\\
& = &
\bigvee_{n\geq m} (e_{n} \ominus e_{m}) &
   \mbox{by Lemma~\ref{lem:ard}~\eqref{lem:ard:minus:oveepres}}
\\
& \leq &
\bigvee_{n\geq m} \ard(e_{n}, e_{m})\cdot 1 \qquad &
   \mbox{by Lemma~\ref{lem:ard}~\eqref{lem:ard:ard}}
\\
& = &
\big(\bigvee_{n \geq m} \ard(e_{n}, e_{m})\big) \cdot 1
\\
& \leq &
\epsilon \cdot 1.
\end{array} \]

\noindent Lemma~\ref{lem:ard}~\eqref{lem:ard:minusleqovee} gives
$\bigvee_{n} e_{n} \leq e_{m} \ovee \epsilon\cdot 1$, and in
particular $\frac{1}{2}\cdot(\bigvee_{n} e_{n}) \leq \frac{1}{2}\cdot
e_{m} \ovee \frac{\epsilon}{2}\cdot 1$. Hence $\ard(\bigvee_{n} e_{n},
e_{m}) \leq \epsilon$, so that $\lim_{m} e_{m} = \bigvee_{n}
e_{n}$. \QED
\end{enumerate}
\end{myproof}

With this information about distances and joins and their relation in
effect modules we return to our main examples from
Lemma~\ref{lem:predfun}. We describe the Archimedean metrics in these
cases in more detail, and discover that we can describe them also as
`validity' metrics, but in dual form: here they involve joins over
states, and not over predicates like in Section~\ref{sec:states}.

\begin{proposition}
\label{prop:preddist}
\begin{enumerate}
\item \label{prop:preddist:Dst} Let $X$ be an arbitrary set. The
  Archimedean metric $\ard$ induced on the effect module $[0,1]^{X}$
  of fuzzy predicates on $X$ is the supremum
  metric~\eqref{eqn:supmet}, as observed
  in~\cite{JacobsM12b,JacobsMF16}. But this metric can alternatively
  be described via validities, as (in the last equation) in:
\[ \begin{array}{rcccccl}
\ard(p,q)
& = &
\spd(p,q)
& \smash{\stackrel{\eqref{eqn:supmet}}{=}} &
\displaystyle\bigvee_{x\in X} \big|p(x) - q(x)\big|
& = &
\displaystyle\bigvee_{\omega\in\Dst(X)} 
   \big|\,\omega\models p - \omega\models q\,\big|.
\end{array} \]

\item \label{prop:preddist:vNA} Let $\mathscr{A}$ be a von Neumann
  algebra. The Archimedean metric $\ard$ on the effect module
  $[0,1]_{\mathscr{A}}$ of effects of $\mathscr{A}$ is the distance
  induced by the norm $\|-\|$ of $\mathscr{A}$. Moreover, this
  distance can be described as on the right below.
\[ \begin{array}{rcccl}
\ard(e,d)
& = &
\big\|\,e-d\,\big\|
& = &
\displaystyle\bigvee_{\omega\colon\mathscr{A}\rightarrow\C}
   \big|\,\omega\models e - \omega\models d\,\big|.
\end{array} \]
\end{enumerate}
\end{proposition}

\begin{myproof}
\begin{enumerate}
\item Let $p,q\in [0,1]^{X}$. We abbreviate $s \coloneqq \bigvee_{x}
  |p(x) - q(x)|$ and $t \coloneqq \bigvee_{\omega} |\omega\models p -
  \omega\models q|$. First note that for $x\in X$ the unit (or
  `Dirac') distribution $\eta(x) = 1\ket{x}$ satisfies $\eta(x)
  \models p = p(x)$. This yields $s \leq t$. The converse inequality
  $t \leq s$ follows from:
\[ \begin{array}{rcl}
t
\hspace*{\arraycolsep}=\hspace*{\arraycolsep}
\bigvee_{\omega} \big|\,\omega\models p - \omega\models q\,\big|
& = &
\bigvee_{\omega} \big|\,\sum_{x}\omega(x)\cdot p(x) - 
   \sum_{x} \omega(x)\cdot q(x)\,\big|
\\
& \leq &
\bigvee_{\omega} \sum_{x} \omega(x) \cdot \big| p(x) - q(x)\big|
\\
& \leq &
\bigvee_{\omega} \sum_{x} \omega(x) \cdot s
\\
& = &
\bigvee_{\omega} (\sum_{x} \omega(x)) \cdot s
\\
& = &
s.
\end{array} \]

\item From~\cite[Cor.~4.3.10]{Kadison2015} we see that for a
  self-adjoint element $a\in\mathscr{A}$ we have $\|a\| =
  \bigvee_{\omega} |\omega(a)|$, where $\omega$ ranges over (normal)
  states $\mathscr{A}\rightarrow\C$. Thus:
\[ \begin{array}{rccclcl}
\ard(e,d)
& = &
\big\|\,e-d\,\big\|
& = &
\bigvee_{\omega}\big|\,\omega(e-d)\,\big|
\\
&& & = &
\bigvee_{\omega}\big|\,\omega(e)-\omega(d)\,\big|
& = &
\bigvee_{\omega}\big|\,\omega\models e-\omega \models d\,\big|.
\end{array} \eqno{\QEDbox} \]
\end{enumerate}
\end{myproof}

\section{State-and-effect triangles}\label{sec:triangles}

In this section the results from the two previous sections are
combined. This will happen via the adjunction $\op{\EMod}
\rightleftarrows \Conv$ between effect modules and convex sets
from~\cite{Jacobs10e}. This adjunction is restricted by imposing
completeness requirements on both sides. Then it is shown how our
standard examples give rise to commuting state-and-effect triangles
with full and faithful state and predicate functors.

Recall from Section~\ref{sec:states} that we write $\ConvMet$ for the
category of convex metric spaces, and $\ConvCMet$ for the subcategory
of convex \emph{complete} metric spaces.

\begin{lemma}
\label{lem:emodconvadj}
The adjunction from~\cite{Jacobs10e} on the left below restricts
to the adjunction on the right.
\begin{equation}
\label{diag:emodconvadj}
\vcenter{\xymatrix@C-0.5pc{
\op{\EMod}\ar@/^2ex/[rr] & \top & 
   \Conv\quad\ar@/^1.5ex/[ll]
&  &
\op{\DcEMod}\ar@/^2ex/[rr]& \top & 
   \ConvCMet\ar@/^1.5ex/[ll] 
}}
\end{equation}

\noindent All functors are given by `homming into $[0,1]$'.
\end{lemma}

\begin{myproof}
The proof boils down to two points:
\begin{enumerate}
\item \label{point1:emodconvadj} For a directed complete effect module
  $E$, the convex set $\DcEMod\big(E, [0,1]\big)$ is a (convex)
  complete metric space.

\item \label{point2:emodconvadj} For a convex complete metric space
  $X$, the effect module $\ConvCMet\big(X, [0,1]\big)$ is directed
  complete.
\end{enumerate}

\noindent As to point~\eqref{point1:emodconvadj}, let $E$ be a
directed complete effect module. The homset $\DcEMod(E,[0,1])$ carries
the supremum metric~\eqref{eqn:supmet}. This metric is complete with
pointwise limits: $(\lim h_{n})(e) = \lim h_{n}(e)$. It is easy to see
that such a limit map $\lim h_{n}$ preserves sums $\ovee$ and scalar
multiplication.  Hence it is a map of effect modules, and thus
automatically a non-expansive (and continuous) function.  In order to
see that it is also Scott-continuous, let $(e_{i})$ be directed
collection of elements in $E$. Writing $h = \lim h_{n}$, with each
$h_n$ Scott-continuous, we have to prove $h(\bigvee e_{i}) = \bigvee
h(e_{i})$. This works as follows. For each $n$ and $j$ we have:
\[ \begin{array}{rcl}
\big| h(\bigvee_{i} e_{i}) - \bigvee_{i} h(e_{i}) \big|
& \leq &
\big| h(\bigvee_{i} e_{i}) - h_{n}(\bigvee_{i} e_{i}) \big| + 
   \big| \bigvee_i h_{n}(e_{i}) - h_{n}(e_{j}) \big| 
\\
& & \qquad + \;
\big| h_{n}(e_{j}) - h(e_{j}) \big| + \big| h(e_{j}) - \bigvee_{i} h(e_{i}) \big|
\\
& \leq &
\spd(h, h_{n}) + \big| \bigvee_{i} h_{n}(e_{i}) - h_{n}(e_{j}) \big| 
+ \spd(h, h_{n}) + \big| h(e_{j}) - \bigvee_{i} h( e_{i}) \big|.
\end{array} \]

\noindent By choosing $n$ suitably large, the two $\spd$ distances can
be made arbitrarily small.  Having fixed~$n$, the term~$\left|
\bigvee_i h_n(e_i)-h_n(e_j) \right|$ can be made arbitrary small too
by choosing~$j$ suitably large, since the directed net
$(\,h_n(e_i)\,)_i$ in~$[0,1]$ converges to its supremum $\bigvee_i
h_n(e_i)$.  Since the final term $\left|h(e_{j})-\bigvee_i h(e_i)\right|$
vanishes too as~$j$ increases we see that~$\left|h(\bigvee_i e_i) -
\bigvee_i h(e_i)\right|=0$, and so~$h(\bigvee_i e_i) = \bigvee_i
h(e_i)$.

The homset $\DcEMod\big(E, [0,1]\big)$ also has a convex structure,
given by the map:
$$\xymatrix{
\Dst\Big(\DcEMod\big(E, [0,1]\big)\Big)\ar[r]^-{\alpha} & 
   \DcEMod\big(E, [0,1]\big)
\quad\mbox{with}\quad
{\begin{array}{rcl}
\alpha(\omega)(e)
& = &
\sum_{h} \omega(h)\cdot h(e),
\end{array}}
}$$

\noindent where $h$ ranges over $\DcEMod\big(E,[0,1]\big)$. Notice
that each element $e\in E$ gives rise to a non-expansive predicate
$\evmap_{e} \colon \DcEMod(E,[0,1]) \rightarrow [0,1]$ via
$\evmap_{e}(h) = h(e)$. It satisfies for $\omega \in
\Dst\big(\DcEMod(E, [0,1])\big)$,
\[ \begin{array}{rcccccl}
\omega\models \evmap_{e}
& = &
\sum_{h} \omega(h)\cdot \evmap_{e}(h)
& = &
\sum_{h} \omega(h) \cdot h(e)
& = &
\alpha(\omega)(e).
\end{array} \]

\noindent Now we can show that the algebra map $\alpha$ on
$\DcEMod\big(E,[0,1]\big)$ is non-expansive, using
the Kantorovich metric~\eqref{eqn:kantorovich} on distributions:
\[ \begin{array}{rcl}
\spd\big(\alpha(\omega_{1}), \alpha(\omega_{2})\big)
\hspace*{\arraycolsep}\smash{\stackrel{\eqref{eqn:supmet}}{=}}
   \hspace*{\arraycolsep}
{\displaystyle\bigvee\!}_{e}\,
   \big|\,\alpha(\omega_{1})(e) - \alpha(\omega_{2})(e)\,\big|
& = &
{\displaystyle\bigvee\!}_{e}\,
   \big|\,\omega_{1}\models\evmap_{e} - \omega_{2}\models\evmap_{e}\,\big|
\\
& \leq &
{\displaystyle\bigvee\!}_{p}\,
   \big|\,\omega_{1}\models p - \omega_{2}\models p\,\big|
\\
& \smash{\stackrel{\eqref{eqn:kantorovich}}{=}} &
\kvd(\omega_{1}, \omega_{2}).
\end{array} \]

\noindent Each map $f\colon E \rightarrow D$ in $\EMod$ gives an
affine map $(-) \after f \colon \Hom(D, [0,1]) \rightarrow \Hom(E,
[0,1])$ in $\Conv$; it is easy to show that it is also non-expansive.

\auxproof{
This pre-composition map is also non-expansive, since for effect
module maps $h,k \colon D \rightarrow [0,1]$,
$$\begin{array}[b]{rcl}
\spd\big(h \after f, k\after f\big)
& \smash{\stackrel{\eqref{eqn:supmet}}{=}} &
{\displaystyle\bigvee\!}_{e\in E}\, \big|\,h(f(e)) - k(f(e))\,\big|
\\
& \leq &
{\displaystyle\bigvee\!}_{d\in D}\, \big|\,h(d) - k(d)\,\big|
\\
& \smash{\stackrel{\eqref{eqn:supmet}}{=}} &
\spd\big(h, k\big).
\end{array}$$
}

For point~\eqref{point2:emodconvadj} we have to prove that for each
convex complete metric space $X$ the set $\ConvCMet(X,[0,1])$ of
affine non-expansive maps is a directed complete effect module. We
concentrate on directed completeness, since the effect module
structure is standard, see~\cite{Jacobs10e}. Hence let $(p_{i})$ be a
directed collection of non-expansive affine maps $p_{i}\colon X
\rightarrow [0,1]$. We take $p = \bigvee_{i} p_{i}$ pointwise. This
map is affine since affine sums are by definition finite, so that
they commute with \emph{directed} joins:
\[ \begin{array}{rcl}
p(\sum_{n} r_{n}\ket{x_{n}})
\hspace*{\arraycolsep}=\hspace*{\arraycolsep}
\big(\bigvee_{i} p_{i}\big)(\sum_{n} r_{n}\ket{x_{n}})
& = &
\bigvee_{i} p_{i}(\sum_{n} r_{n}\ket{x_{n}})
\\
& = &
\bigvee_{i} \sum_{n} r_{n} \cdot p_{i}(x_{n})
\\
& = &
\sum_{n} r_{n} \cdot \big(\bigvee_{i} p_{i}(x_{n})\big)
\hspace*{\arraycolsep}=\hspace*{\arraycolsep}
\sum_{n} r_{n} \cdot p(x_{n}).
\end{array} \]

\noindent It is not hard to see that $p$ is non-expansive. \QED

\auxproof{
\[ \begin{array}{rcl}
\big| p(x) - p(y) \big|
& = &
\big| \bigvee_{i} p_{i}(x) - \bigvee_{i} p_{i}(y) \big|
\\
& \leq &
\bigvee_{i} \big| p_{i}(x) - p_{i}(y) \big|
\\
& \leq &
\bigvee_{i} d_{X}(x, y)
\\
& = &
d_{X}(x, y).
\end{array} \]

\noindent For the first inequality, let us assume without loss of
generality, that $\bigvee_{i} p_{i}(x) \geq \bigvee_{i} p_{i}(y)$. Then:
\[ \begin{array}{rcccccl}
\big| \bigvee_{i} p_{i}(x) - \bigvee_{i} p_{i}(y) \big|
& = &
\bigvee_{i} p_{i}(x) - \bigvee_{i} p_{i}(y)
& = &
\bigvee_{i} \big(p_{i}(x) - p_{i}(y)\big)
& \leq &
\bigvee_{i} \big|p_{i}(x) - p_{i}(y)\big|.
\end{array} \]
}
\end{myproof}

The next two results summarise our main concrete findings.

\begin{proposition}
\label{prop:triangle:Dst}
The Kleisli subcategory $\Klf(\Dst)$, with finite sets only, of the
distribution monad $\Dst$ on $\Sets$ gives rise to a triangle as
below, in which the two up-going functors are full and faithful and
make the two corresponding triangles commute up-to natural isomorphism.
\medskip
\[ \vcenter{\xymatrix@C-1pc{
\op{\DcEMod}\ar@/^2ex/[rr] & \top & 
   \ConvCMet\ar@/^1.5ex/[ll] 
\\
& \Klf(\Dst)\ar[ul]^{\Hom(-,2)=\Pred\quad}\ar[ur]_{\quad\Stat = \Hom(1,-)} &
}} \]
\end{proposition}

We briefly explain the functor $\Pred = \Hom(-,2) \colon \Klf(\Dst)
\rightarrow \op{\EMod}$. Since $\Dst(2) \cong [0,1]$ we get $\Pred(X)
= \Hom(X,2) = \Sets(X,\Dst(2)) = \Sets(X,[0,1]) = [0,1]^X$.

\begin{myproof}
We use the full and faithful predicate functor $\Pred = [0,1]^{(-)}
\colon \Klf(\Dst) \rightarrow \op{\DcEMod}$ from
Lemma~\ref{lem:predfun}~\eqref{lem:predfun:Dst}. The states functor
$\Stat \colon \Klf(\Dst) \rightarrow \Conv = \EM(\Dst)$ is the full
and faithful Kleisli extension functor, restricted to finite sets. The
functor restricts to \emph{metric} spaces $\ConvMet \hookrightarrow
\Conv$ by Lemma~\ref{lem:kantorovich} and to \emph{complete} spaces
$\ConvCMet \hookrightarrow \ConvMet$ by
Lemma~\ref{lem:discretecompleteness}. We need to check that the two
triangles commute.

In one direction we have, for a finite set $X$,
\[ \begin{array}{rcl}
\Big(\DcEMod\big(-,[0,1]\big) \after \Pred\Big)(X)
& = &
\DcEMod\big([0,1]^{X}, [0,1])
\\
& \cong &
\Klf(\Dst)\big(1, X\big)  \qquad \mbox{since $\Pred$ is full \& faithful}
\\
& \cong &
\Dst(X)
\\
& = &
\Stat(X)
\end{array} \]

\noindent In the other direction:
\[ \begin{array}[b]{rcl}
\Big(\ConvCMet\big(-, [0,1]\big) \after \Stat\Big)(X)
& = &
\ConvMet\big(\Dst(X), [0,1]\big)
\\
& \cong &
\Met(X, [0,1]) \quad \mbox{using $X$ with discrete metric}
\\
& = &
\Sets(X, [0,1]) 
\\
& = &
\Pred(X).
\end{array} \eqno{\QEDbox} \]
\end{myproof}

The description, in the above triangle, of the predicate and state
functors via homsets $\Hom(-,2)$ and $\Hom(1,-)$ comes from effectus
theory~\cite{Jacobs15d,ChoJWW15b}. It also applies to von Neumann
algebras, when we use their category in opposite form, as
$\op{\vNA}$. For instance, the initial object in $\vNA$ is the algebra
$\C$ of complex numbers; it forms the final object $1$ in
$\op{\vNA}$. Thus, a map $1 \rightarrow \mathscr{A}$ in $\op{\vNA}$ is
a state $\mathscr{A} \rightarrow \C$, as we have described before. In
a similar way one can check that maps $\mathscr{A} \rightarrow 2 =
1+1$ in $\op{\vNA}$ correspond to effects in the unit interval
$[0,1]_{\mathscr{A}}$, see below, or~\cite{Jacobs15d} for details.

\begin{proposition}
\label{prop:triangle:vNA}
The opposite of the category $\vNA$ of von Neumann algebras fits in a
triangle as below, in which the predicate and state functors are full
and faithful and make the triangles commute up-to natural isomorphism.
\medskip
\[ \vcenter{\xymatrix@C-1pc{
\op{\DcEMod}\ar@/^2ex/[rr] & \top & 
   \ConvCMet\ar@/^1.5ex/[ll] 
\\
& \op{\vNA}\ar[ul]^{\Hom(-,2)=\Pred\quad}\ar[ur]_{\quad\Stat = \Hom(1,-)} &
}} \]
\end{proposition}

The predicate functor $\Pred = \Hom(-,2) \colon \op{\vNA} \rightarrow
\op{\EMod}$ can be described via maps into $2$ in the following
way. The object $2 = 1+1$ is formed in $\op{\vNA}$. Hence it is
$0\times 0$ in $\vNA$, where the initial object $0$ is the algebra
$\C$ of complex numbers. One then needs to check that $\Hom(\C^{2},
\mathscr{A}) \cong [0,1]_{\mathscr{A}}$ for a von Neumann algebra
$\mathscr{A}$ via $f\mapsto f(1,0)$,
which is easy, and left to the reader.
In a similar way the maps in $\Hom(1,\mathscr{A})$ are
the maps of von Neumann algebras $\mathscr{A} \rightarrow \C$. These
are the states, as used before.

\begin{myproof}
In Lemma~\ref{lem:predfun}~\eqref{lem:predfun:vNA} we have seen that
the predicate functor $\Pred = [0,1]_{(-)} \colon \op{\vNA}
\rightarrow \op{\DcEMod}$ is full and faithful. For convenience
we abbreviate $\mathcal{F} = \ConvCMet(-, [0,1])$ and $\mathcal{G}
= \DcEMod(-, [0,1])$ so that $\mathcal{F} \dashv \mathcal{G}$ at
the top of the above triangle.

Starting from the predicate functor $\Pred$ the above triangle
commutes, since $\Pred$ is full and faithful:
\[ \begin{array}{rcl}
\mathcal{G}\Pred(\mathscr{A})
\hspace*{\arraycolsep}=\hspace*{\arraycolsep}
\DcEMod\Big(\Pred(\mathscr{A}), [0,1]\Big)
& = &
\DcEMod\Big(\Pred(\mathscr{A}), \Pred(\C)\Big)
\\
& \cong &
\vNA\big(\mathscr{A}, \C\big)
\\
& = &
\Stat(\mathscr{A}).
\end{array} \]

Commutation of the second triangle is less obvious. It relies on 
    some facts concerning the
    linear combinations of normal states
on~$\mathscr{A}$, which form a closed linear subspace~$\mathscr{A}_*$ of the
continuous dual~$\mathscr{A}^*
    = \set{f\colon \scrA\to\C}{f\text{ is bounded and linear}}$ 
    of~$\mathscr{A}$ (see~\eg~A90, A91, and A92 of~\cite{AlfsenGeom}.) 
    This ``pre-dual'' $\mathscr{A}_*$ of~$\mathscr{A}$ determines the
order and norm of~$\mathscr{A}$ in the sense that the map $a\mapsto
\hat{a}\colon\mathscr{A}\to(\mathscr{A}_*)^*$ which
sends~$a\in\mathscr{A}$ to the bounded
functional~$\hat{a}\colon\mathscr{A}_*\to\C$ given
by~$\hat{a}(\varphi) = \varphi(a)$ is a linear
isomorphism~$\mathscr{A}\to(\mathscr{A}_*)^*$ that preserves (and
reflects) both the norm and the order (see~A94 of~\cite{AlfsenGeom}). 
    Restricted to effects, we get a
natural isomorphism $\Pred = [0,1]_{(-)}\Rightarrow [0,1]_{((-)_*)^*}$.
Since a bounded linear functional $f\colon \mathscr{A}_*\to\C$ on the
pre-dual~$\mathscr{A}_*$ of a von Neumann algebra~$\mathscr{A}$ is
completely determined by its action on the states of~$\mathscr{A}$,
and this action is affine, contractive and maps into~$[0,1]$ when~$f$
is from~$[0,1]_{(\mathscr{A}_*)^*}$, restricting such~$f$ to states
gives a natural isomorphism $[0,1]_{((-)_*)^*} \to
\ConvCMet(\Stat(-),[0,1])$.  Composing this with the natural
isomorphism mentioned before we get a natural isomorphism
$\Pred\Rightarrow\mathcal{F}\Stat$.

With these isomorphisms $\mathcal{G}\Pred \cong \Stat$ and $\Pred \cong
\mathcal{F}\Stat$ in place we can show that the functor $\Stat
\colon \op{\vNA} \rightarrow \ConvCMet$ is full and faithful, since
for two von Neumann algebras $\mathscr{A}$ and $\mathscr{B}$ we have:
\[ \begin{array}[b]{rcl}
\ConvCMet\Big(\Stat(\mathscr{A}), \Stat(\mathscr{B})\Big)
& \cong &
\ConvCMet\Big(\Stat(\mathscr{A}), \mathcal{G}\Pred(\mathscr{B})\Big)
\\
& \cong &
\op{\DcEMod}\Big(\mathcal{F}\Stat(\mathscr{A}), \Pred(\mathscr{B})\Big)
\\
& \cong &
\op{\DcEMod}\Big(\Pred(\mathscr{A}), \Pred(\mathscr{B})\Big)
\\
& \cong &
\op{\vNA}\Big(\mathscr{A}, \mathscr{B}\Big).
\end{array} \eqno{\QEDbox} \]
\end{myproof}

\section{Concluding remarks}\label{sec:conclusions}

In this paper we have given a systematic unifying description of
metrics on states and predicates from the perspective of the duality
between state transformers and predicate transformers, notably in
state-and-effect triangles. This unifying perspective is most
prominent in the use of `validity' metrics, both on states (via joins
over predicates) and on predicates (via joins over states).

We have concentrated on the \emph{discrete} version of classical
probability and on quantum probability. What about \emph{continuous}
classical probability? Most of it has already been done
in~\cite{ChaputDPP14}, see also~\cite{ClercDDG17}, albeit in slightly
different form, using $\omega$-complete ordered cones instead of
directed complete effect modules, together with a `cone duality'
result of the form $\Hom(L^{+}_{p}(X,\mu), \nnR) \cong
L^{+}_{q}(X,\mu)$ when $\frac{1}{p} + \frac{1}{q} = 1$; here, $X$ is a
measurable space with measure $\mu$. In the language of triangles,
this duality corresponds to commutation of the triangles, as in the
above Propositions~\ref{prop:triangle:Dst}
and~\ref{prop:triangle:vNA}. In a next step, as in~\cite{ClercDDG17},
a category of `kernels' can be formed, as comma category $(1\downarrow
\cat{B})$ of the base category $\cat{B}$ that we use in triangles.
For instance, the comma category $(1\downarrow\Kl(\Dst))$ contains
distributions as objects, and distribution preserving maps between
them. They can be used to define Bayesian inversion in the form of a
dagger functor on such a comma category, see notable~\cite{ClercDDG17}
--- and~\cite{ChoJ19} for a wider perspective on inversion and
disintegration.

\bibliographystyle{plain}

{ \appendix

\section{Missing proofs from Section~\ref{sec:states}}

\begin{appendixproof}{of Proposition~\ref{prop:totvardist}}
Let $\omega_{1},\omega_{2}\in\Dst(X)$ be two discrete probability
distributions on the same set $X$. Recall from~\eqref{eqn:totvardist}
that by definition: $\tvd\big(\omega_{1}, \omega_{2}\big) =
\frac{1}{2}\sum_{x\in X}\big|\,\omega_{1}(x) -
\omega_{2}(x)\,\big|$. We will prove the two inequalities labeled
$(a)$ and $(b)$ in:
\[ \begin{array}{rcccccl}
\tvd\big(\omega_{1}, \omega_{2}\big)
& \stackrel{(a)}{\leq} &
\displaystyle\max\limits_{U\subseteq X}\,
   \omega_{1}\models \indic{U} - \omega_{2}\models \indic{U}
& \leq &
\!\displaystyle\bigvee_{p\in[0,1]^{X}} 
   \Big|\,\omega_{1}\models p - \omega_{2}\models p\,\Big|
& \stackrel{(b)}{\leq} &
\tvd\big(\omega_{1}, \omega_{2}\big).
\end{array} \]

\noindent This proves Proposition~\ref{prop:totvardist} since the
inequality in the middle is trivial.

We start with some preparatory definitions.  Let $U\subseteq X$ be an
arbitrary subset. Recall that we write $\omega_{i}(U) = \sum_{x\in
	U}\omega_{i}(x) = (\omega\models\indic{U})$. We partition $U$ in three
disjoint parts, and take the relevant sums:
\[ \left\{\begin{array}{rcl}
U_{>}
& = &
\setin{x}{U}{\omega_{1}(x)>\omega_{2}(x)}
\\
U_{=}
& = &
\setin{x}{U}{\omega_{1}(x)=\omega_{2}(x)}
\\
U_{<}
& = &
\setin{x}{U}{\omega_{1}(x)<\omega_{2}(x)}
\end{array}\right.
\qquad\qquad
\left\{\begin{array}{rcccl}
\upsum{U}
& = &
\omega_{1}(U_{>}) - \omega_{2}(U_{>})
& \geq &
0
\\
\downsum{U}
& = &
\omega_{2}(U_{<}) - \omega_{1}(U_{<})
& \geq &
0.
\end{array}\right. \]

\noindent We use this notation in particular for $U = X$. In that
case we can use:
\[ \begin{array}{rcccl}
1
& = &
\omega_{1}(X)
& = &
\omega_{1}(X_{>}) + \omega_{1}(X_{=}) + \omega_{1}(X_{<}) 
\\
1
& = &
\omega_{2}(X)
& = &
\omega_{2}(X_{>}) + \omega_{2}(X_{=}) + \omega_{2}(X_{<}) 
\end{array} \]

\noindent Hence by subtraction we obtain, since $\omega_{1}(X_{=}) =
\omega_{2}(X_{=})$,
\[ \begin{array}{rcl}
0
& = &
\big(\omega_{1}(X_{>}) - \omega_{2}(X_{>})\big) +
   \big(\omega_{1}(X_{<}) - \omega_{2}(X_{<})\big)
\end{array} \]

\noindent That is,
\[ \begin{array}{rcccccl}
\upsum{X}
& = &
\omega_{1}(X_{>}) - \omega_{2}(X_{>})
& = &
\omega_{2}(X_{<}) - \omega_{1}(X_{<})
& = &
\downsum{X}.
\end{array} \]

\noindent As a result:
\begin{equation}
\label{eqn:totvardistaux}
\begin{array}{rcl}
\tvd\big(\omega_{1}, \omega_{2}\big)
& = &
\frac{1}{2}\,\sum_{x\in X} \big|\,\omega_{1}(x) - \omega_{2}(x)\big|
\\
& = &
\frac{1}{2}\Big(\sum_{x\in X_{>}} \big(\,\omega_{1}(x) - \omega_{2}(x)\big) +
   \sum_{x\in X_{<}} \big(\,\omega_{2}(x) - \omega_{1}(x)\big)\Big)
\\[1mm] 
& = &
\frac{1}{2}\Big(\big(\,\omega_{1}(X_{>}) - \omega_{2}(X_{>})\big) +
   \big(\,\omega_{2}(X_{<}) - \omega_{1}(X_{<})\big)\Big)
\\
& = &
\frac{1}{2}\big(\upsum{X} \,+\; \downsum{X}\!\big)
\\
& = &
\upsum{X}
\end{array}
\end{equation}

\noindent We have prepared the ground for proving the above inequalities
$(a)$ and $(b)$.
\begin{enumerate}
\item[$(a)$] We will see that the above maximum is actually reached for
  the subset $U = X_{>}$, first of all because:
\[ \begin{array}{rcl}
\tvd\big(\omega_{1}, \omega_{2}\big)
\hspace*{\arraycolsep}\smash{\stackrel{\eqref{eqn:totvardistaux}}{=}}
   \hspace*{\arraycolsep}
\upsum{X}
\hspace*{\arraycolsep}=\hspace*{\arraycolsep}
\omega_{1}(X_{>}) - \omega_{2}(X_{>})
& = &
\omega_{1}\models\indic{X_{>}} - \omega_{2}\models\indic{X_{>}}
\\
& \leq &
\displaystyle\max\limits_{U\subseteq X}\,
   \omega_{1}\models \indic{U} - \omega_{2}\models \indic{U}.
\end{array} \]

\item[$(b)$] Let $p\in[0,1]^{X}$ be an arbitrary predicate. We write
  $\indic{U}\andthen p$ for the pointwise product predicate, with:
  $\big(\indic{U}\andthen p\big) = \indic{U}(x)\cdot p(x)$, which is
  $p(x)$ if $x\in U$ and $0$ otherwise. Then:
\[ \begin{array}[b]{rcl}
\lefteqn{\big|\omega_{1}\models p - \omega_{2}\models p\big|}
\\
& = &
\Big|\big(\omega_{1}\models\indic{X_{>}}\andthen p \;+\;
   \omega_{1}\models\indic{X_{=}}\andthen p \;+\;
   \omega_{1}\models\indic{X_{<}}\andthen p\big)
\\
& & \quad -\;
\big(\omega_{2}\models\indic{X_{>}}\andthen p \;+\;
   \omega_{2}\models\indic{X_{=}}\andthen p \;+\;
   \omega_{2}\models\indic{X_{<}}\andthen p\big)\Big|
\\
& = &
\Big|\big(\omega_{1}\models\indic{X_{>}}\andthen p \,-\,
          \omega_{2}\models\indic{X_{>}}\andthen p\big) \;-\;
   \big(\omega_{2}\models\indic{X_{<}}\andthen p \,-\,
   \omega_{1}\models\indic{X_{<}}\andthen p\big)\Big|
\\
& = &
\left\{\begin{array}{l}
\big(\omega_{1}\models\indic{X_{>}}\andthen p \,-\,
          \omega_{2}\models\indic{X_{>}}\andthen p\big) \;-\;
   \big(\omega_{2}\models\indic{X_{<}}\andthen p \,-\,
   \omega_{1}\models\indic{X_{<}}\andthen p\big) \\
\quad \mbox{if } 
   \big(\omega_{1}\models\indic{X_{>}}\andthen p \,-\,
          \omega_{2}\models\indic{X_{>}}\andthen p\big) 
   \,\stackrel{(*)}{\geq}\,
   \big(\omega_{2}\models\indic{X_{<}}\andthen p \,-\,
   \omega_{1}\models\indic{X_{<}}\andthen p\big)
\\
\big(\omega_{2}\models\indic{X_{<}}\andthen p \,-\,
   \omega_{1}\models\indic{X_{<}}\andthen p\big) \;-\;
   \big(\omega_{1}\models\indic{X_{>}}\andthen p \,-\,
      \omega_{2}\models\indic{X_{>}}\andthen p\big) 
    \\
\qquad \mbox{otherwise}
\end{array}\right.
\\
& \leq &
\left\{\begin{array}{ll}
\omega_{1}\models\indic{X_{>}}\andthen p \,-\,
          \omega_{2}\models\indic{X_{>}}\andthen p
   \quad & \mbox{if }(*)
\\
\omega_{2}\models\indic{X_{<}}\andthen p \,-\,
   \omega_{1}\models\indic{X_{<}}\andthen p & \mbox{otherwise}
\end{array}\right.
\\
& = &
\left\{\begin{array}{ll}
\sum_{x\in X_{>}} (\omega_{1}(x)-\omega_{2}(x))\cdot p(x) 
   \quad & \mbox{if }(*)
\\
\sum_{x\in X_{<}} (\omega_{2}(x)-\omega_{1}(x))\cdot p(x) & \mbox{otherwise}
\end{array}\right.
\\
& \leq &
\left\{\begin{array}{ll}
\sum_{x\in X_{>}} \omega_{1}(x)-\omega_{2}(x) \quad & \mbox{if }(*)
\\
\sum_{x\in X_{<}} \omega_{2}(x)-\omega_{1}(x) & \mbox{otherwise}
\end{array}\right.
\\
& = &
\left\{\begin{array}{ll}
\upsum{X} & \mbox{if }(*)
\\
\downsum{X} \;= \upsum{X} \quad  & \mbox{otherwise}
\end{array}\right.
\\
& = &
\upsum{X}
\\
& \smash{\stackrel{\eqref{eqn:totvardistaux}}{=}} &
\tvd\big(\omega_{1}, \omega_{2}\big).
\end{array} \eqno{\QEDbox} \]
\end{enumerate}
\end{appendixproof}

\medskip

\begin{appendixproof}{of Proposition~\ref{prop:trace}}
Let $\varrho_{1},\varrho_{2}$ be two states (density operators) of a
Hilbert space $\mathscr{H}$.  The trick is to split the trace-class
operator $\varrho \coloneqq \varrho_1-\varrho_2$ into its positive and
negative parts: we have~$\varrho=\varrho_+-\varrho_-$,
where~$\varrho_+,\varrho_-\colon \mathscr{H}\to\mathscr{H}$ are
positive operators with $\varrho_+\varrho_-=0$
and~$\left|\varrho\right|=\varrho_++\varrho_-$,
see~\cite[Cor~2.15]{Alfsen2012}.  Note that since
$\varrho_+,\varrho_-\leq \left|\varrho\right|$ the operators
$\varrho_+$ and~$\varrho_-$ are trace-class as well.  The key is to
note that $\tr(\varrho_+)-\tr(\varrho_-) =
\tr(\varrho)=\tr(\varrho_1)-\tr(\varrho_2)=1-1=0$, so
that~$\tr(\varrho_+)=\tr(\varrho_-)$. Hence:
\[ \begin{array}{rcccccccl}
\trd(\varrho_1,\varrho_2)
& \smash{\stackrel{\eqref{eqn:tracedist}}{=}} &
\frac{1}{2}\tr(\left|\varrho\right|)
& = &
\frac{1}{2}\big(\tr(\varrho_+)+\tr(\varrho_-)\big)
& = &
\tr(\varrho_+)
& = &
\tr(\varrho_-).
\end{array} \]

\noindent Now, given an effect~$p$ on~$\mathscr{H}$ we have $\varrho_1
\models p - \varrho_2\models p = \tr(\varrho_{1}\, p) -
\tr(\varrho_{2}\, p) = \tr(\varrho\, p) =\tr(\varrho_+\, p) -
\tr(\varrho_-\, p) \leq \tr(\varrho_+\, p) \leq \tr(\varrho_+) =
\trd(\varrho_1,\varrho_2)$, using $p\leq\idmap$.
    (Here we used that~$\tr(\varrho_-\,p)\geq 0$
    by A87 of~\cite{AlfsenGeom},
    because~$\varrho_-\geq 0$ and~$p\geq 0$.)
    Since similarly
$\varrho_2 \models p - \varrho_1 \models p \leq
\trd(\varrho_1,\varrho_2)$, we get:
\[ \begin{array}{rcl}
\displaystyle\bigvee_{p \in \Ef(\mathscr{H}) }
  \big|\,\varrho_2 \models p - \varrho_1 \models p\,\big|
& \leq &
\trd(\varrho_1,\varrho_2).
\end{array} \]

\noindent The only thing that remains to be shown is that there is a
projection~$s$ on~$\mathscr{H}$ with~$\varrho_1\models s -
\varrho_2\models s = \trd(\varrho_1,\varrho_2)$.  It turns out that we
need to pick the least projection~$s$ in~$\mathscr{B}(\mathscr{H})$
with $\varrho_+ s=\varrho_+$ (which exists, see
\textit{e.g.}~\cite[Defn~2.107]{Alfsen2012}).  If~$t$ denotes the
least projection with $\varrho_- t=\varrho_-$ then one can prove
    that~$ts=0$ (see \eg~59IV1 of~\cite{bram}), so that~$\varrho_-s=\varrho_-ts=0$.  Whence
$\varrho_1\models s - \varrho_2\models s = \tr(\varrho_{1} s) -
\tr(\varrho_{2} s) = \tr(\varrho s) = \tr(\varrho_+s) -
\tr(\varrho_-s) = \tr(\varrho_+)=\trd(\varrho_1,\varrho_2)$. \QED

\end{appendixproof}

\medskip

\begin{appendixproof}{of Proposition~\ref{prop:vlddist}}
Let $\varrho_{1}, \varrho_{2} \colon \mathscr{A} \rightarrow \C$ be
two (normal) states of a von Neumann algebra $\mathscr{A}$ and let~$e\in
[0,1]_\mathscr{A}$ be an arbitrary effect. If we bluntly apply the
definition of the operator norm we only get $\left| \varrho_1 \models
e - \varrho_2\models e\right| = \left|(\varrho_1-\varrho_2)(e)\right|
\leq \opnorm{\varrho_1-\varrho_2}\cdot \|e\| \leq
\opnorm{\varrho_1-\varrho_2}$. The factor~``$\frac{1}{2}$'' from
Proposition~\ref{prop:vlddist} is then missing, so a more subtle
approach is called for.  Writing $\varrho\coloneqq\varrho_1-\varrho_2$
there is by~\cite[Thm~7.4.7]{Kadison2015} a sharp predicate $s\in
[0,1]_A$ such that both~$\varrho_+ \coloneqq \varrho(s(\,\cdot\,)s)$
and~$\varrho_- \coloneqq -\varrho(s^\perp(\,\cdot\,)s^\perp)$ are
positive and normal, and, moreover,
\[ \begin{array}{rclcrcl}
\varrho
& = &
\varrho_+ - \varrho_-
& \qquad\text{and}\qquad &
\opnorm{\varrho}
& = &
\opnorm{\varrho_+} + \opnorm{\varrho_-}.
\end{array} \]

\noindent Further, by~\cite[Thm~4.3.2]{Kadison2015} we
have~$\opnorm{\varrho_1} = \varrho_1(1)$ and~$\opnorm{\varrho_2} =
\varrho_2(1)$.  Then since~$\varrho_1$ and~$\varrho_2$ are states, we
have $\varrho(1)=\varrho_1(1)-\varrho_2(1)=1-1=0$,
so~$\varrho_+(1)-\varrho_-(1)=\varrho(1)=0$, and
thus~$\opnorm{\varrho_+} = \varrho_+(1) = \varrho_-(1) =
\opnorm{\varrho_-}$.  But then, since~$\opnorm{\varrho} =
\opnorm{\varrho_+} + \opnorm{\varrho_-}$, we get:
\[ \begin{array}{rcccl}
\opnorm{\varrho_+}
& = &
\opnorm{\varrho_-}
& = &
\frac{1}{2}\opnorm{\varrho_1-\varrho_2}.
\end{array} \]

\noindent Now, given~$e\in [0,1]_A$ we have $\varrho_1\models
e-\varrho_2\models e = \varrho(e)\leq \varrho_+(e)\leq \varrho_+(1)
\leq \opnorm{\varrho_+} = \frac{1}{2} \opnorm{\varrho_1-\varrho_2}$,
and so~$\bigvee_{e\in [0,1]_{\mathscr{A}}} \varrho_1\models e -
\varrho_2\models e \leq \frac{1}{2} \opnorm{\varrho_1-\varrho_2}$.  By
a similar reasoning, we get $\bigvee_{e\in [0,1]_A} \varrho_2\models e
- \varrho_1\models e \leq \frac{1}{2} \opnorm{\varrho_1-\varrho_2}$,
and so:
\[ \begin{array}{rcl}
\displaystyle\bigvee_{e\in [0,1]_{\mathscr{A}}} 
   \big|\,\varrho_1\models e - \varrho_1\models e\,\big|
& \leq &
\frac{1}{2}\opnorm{\varrho_1-\varrho_2}.
\end{array} \]

\noindent The only real thing left to prove is
that~$\frac{1}{2}\opnorm{\varrho_1-\varrho_2}=
\varrho_1(s)-\varrho_2(s)$, for the above sharp predicate $s$, because
all the equalities in Proposition~\ref{prop:vlddist} follow trivially
from it.  Since~$\varrho_+=\varrho(s(\,\cdot\,)s)$ we have
$\varrho_+(s)=\varrho(s)=\varrho_+(1)=\opnorm{\varrho_+} =
\frac{1}{2}\opnorm{\varrho_1-\varrho_2}$; and
since~$\varrho_-=-\varrho(s^\perp(\,\cdot\,)s^\perp)$ we have
$\varrho_-(s)=-\varrho(s^\perp s s^\perp ) = -\varrho(0) = 0$.  Whence
$\varrho_1(s)-\varrho_2(s) = \varrho(s) = \varrho_+(s)-\varrho_-(s) =
\varrho_+(s) = \frac{1}{2}\opnorm{\varrho_1-\varrho_2}$. \QED
\end{appendixproof}

}

\end{document}